\documentclass[%
prc,
twocolumn,
floatfix,
showpacs,
showkeys,
altaffilletter,
 amsmath,amssymb,
 aps,
]{revtex4-1}

\usepackage{graphicx}
\usepackage{dcolumn}
\usepackage{bm}
\usepackage{epstopdf}
\usepackage{gensymb}
\usepackage{rotating}

\begin{document}


\title{High-spin spectroscopy and shell-model interpretation of the \textit{N}~$<$~126 radium isotopes $^{212}$Ra and $^{213}$Ra}
				
\author{T.~Palazzo}
\author{G.~J.~Lane}
\author{A.~E.~Stuchbery}
\author{A.~J.~Mitchell}
\author{A.~Akber}
\author{M.~S.~M.~Gerathy}
\author{S.~S.~Hota}
\author{T.~Kib{\'e}di}
\author{B.~Q.~Lee}
\author{N.~Palalani}
\author{M.~W.~Reed}
\affiliation{Department of Nuclear Physics, Research School of Physics and Engineering, The Australian National University, Canberra, ACT 2601, Australia}

\begin{abstract}
The level structures of $^{212}$Ra and $^{213}$Ra have been established via time-correlated $\gamma$-ray spectroscopy following the $^{204}$Pb($^{12}$C,4$n$)$^{212}$Ra and $^{204}$Pb($^{13}$C,4$n$)$^{213}$Ra reactions.  In $^{212}$Ra, levels up to $\sim 6.2$~MeV were identified and firm spin-parity assignments were achieved to a $J^{\pi} = 19^+$ isomer with a mean life of 31(3)~ns. For $^{213}$Ra the corresponding values were $\sim 4.5$~MeV in excitation energy and $J^{\pi} = 33/2^+$. Two isomeric states with $J^\pi$~=~23/2$^+$, $\tau$~=~27(3)~ns and $J^\pi$~=~33/2$^+$, $\tau$~=~50(3)~ns were discovered in $^{213}$Ra. The experimental data were compared with semiempirical shell-model calculations, which allowed dominant configurations to be assigned to most of the observed levels.
\end{abstract}



\maketitle

\section{Introduction}
The nuclear shell model \cite{Mayer.PhysRev.74.235} is the foundation on which much of our understanding of atomic nuclei has been built.
Nuclei in the vicinity of doubly magic $^{208}$Pb ($Z$~=~82, $N$~=~126) provide an important testing ground for the validity of the shell model. More recently they have served as a benchmark for studies of the shell structure around neutron-rich $^{132}$Sn, which has become experimentally accessible through advances in experimental capabilities with radioactive ion beams \cite{Coraggio.PhysRevC.80.021305,Gargano2009}. In this paper we focus on the radium isotopes ($Z=88$) below $N=126$, namely $^{213}$Ra ($N$~=~125) and $^{212}$Ra ($N$~=~124). The shell model approach has proved to be applicable to the $N$ = 126 isotones above $Z=82$: $^{210}$Po ($Z$~=~84) \cite{po210}, $^{211}$At ($Z$~=~85) \cite{BAYER2001}, $^{212}$Rn ($Z$~=~86) \cite{rn212}, $^{213}$Fr ($Z$~=~87) \cite{Fr211-12-13} and $^{214}$Ra ($Z$~=~88) \cite{ra214}, at least to moderate spins.
Lifetime measurements and the resulting B($E$2) values obtained for the radium isotopes with $N$~$>$~126 suggest a smooth increase in collectivity as the number of valence neutrons increases \cite{ra215,ra216-7,ra217a}. Knowledge of the radium isotopes near but below $N=126$, particularly $^{213}$Ra and $^{212}$Ra, is more limited. Technical challenges associated with measuring time correlated $\gamma$-ray coincidences across long-lived isomeric states, along with low production cross sections and strong fission competition, have hindered spectroscopic studies of these nuclei to higher spins. Prior knowledge of the decay scheme of $^{213}$Ra has been limited to a single cascade of three $\gamma$-ray transitions below a $J^\pi$~=~17/2$^-$, $\tau$~=~3-ms isomer \cite{raich-ra213}. Spectroscopic data on the neighbouring isotope, $^{212}$Ra, is more extensive, but achieves firm spin assignments only to $J^{\pi} = 13^-$ \cite{ra212}.

Here we report on the experimental extension of the level schemes of $^{213}$Ra and $^{212}$Ra up to $J^{\pi} = 33/2^+$ and $J^{\pi} = 19^+$, respectively. 
The new level schemes are considered within a semiempirical shell-model framework, which allows the assignment of the dominant configuration to many of the observed states. 
Limitations of the calculations as additional nucleons are added to the valence space are also discussed.

\section{Experimental details}

Excited states in $^{213}$Ra and $^{212}$Ra were populated via the $^{204}$Pb($^{13}$C, 4$n$)$^{213}$Ra and $^{204}$Pb($^{12}$C, 4$n$)$^{212}$Ra reactions, with beam energies of 80~MeV and 81~MeV, respectively.
The $^{12,13}$C beams were delivered by the 14UD accelerator of the Heavy Ion Accelerator Facility at the Australian National University and pulsed to $\sim 1$~ns in width separated by 1712~ns. The target was isotopically enriched $^{204}$Pb (99.6$\%$), 5.4-mg/cm$^2$ thick.

Prompt (in-beam) and delayed (out-of-beam) emission of $\gamma$ rays was measured using the CAESAR array of Compton-suppressed, high-purity germanium (HPGe) detectors. CAESAR consists of nine HPGe detectors in a close-packed geometry, six of which are positioned in pairs perpendicular to the beam axis in the vertical plane at angles of $\pm$34$\degree$, $\pm$48$\degree$ and $\pm$82$\degree$. Three HPGe detectors, in addition to two unsuppressed, low-energy-photon spectrometers (LEPS), were positioned approximately in the horizontal plane. Time-correlated $\gamma-\gamma$ coincidence data were collected in list mode. The HPGe and LEPS detectors were calibrated to 0.5-keV and 0.2-keV per channel, respectively.
	
Angular anisotropies were measured via $\gamma$-$\gamma$ coincidence data sorted into two-dimensional matrices, with pairs of detectors at equivalent angles of $\pm$34$\degree$, $\pm$48$\degree$ and $\pm$82$\degree$ recorded on one axis and any measured, coincident $\gamma$ ray in the remaining eight HPGe detectors placed on the other axis. Since only three angle pairs were available, the $A_4$ coefficient was fixed to zero when fitting the measured angular data with the usual expansion in even-order Legendre polynomials. Extracted $A_2$ values still serve as a guide to determine transition multipolarities; however, the limited angle coverage and need to set $A_4=0$ meant that precise transition mixing ratios could not be determined. For the angular momentum alignment expected in heavy-ion fusion-evaporation reactions, and with coincidence gates placed on known $E2$ transitions, pure dipole, quadrupole or octupole transitions typically exhibit an $A_2/A_0$ value of $-0.2$, $+0.28$ or $+0.46$, respectively. In some cases, it was possible to make definitive transition-multipolarity assignments by determining the internal conversion coefficients from $\gamma$-ray intensity balances and comparison with theoretical values \cite{bricc}. Since the probability for internal conversion decreases significantly with increasing transition energy, this method was only applicable to low-energy transitions.
	
Isomeric-level lifetime measurements were made by observing the time of arrival of $\gamma$ rays with respect to the beam pulse, or by evaluation of the time difference between $\gamma$ rays feeding and depopulating the state of interest. In the former case, all levels populated following the $\gamma$ decay of an isomeric state exhibit that lifetime, thereby inhibiting the measurement of shorter-lived isomers located below the long-lived states. Through complementary application of the $\gamma\gamma$ time-difference method, it has been possible to isolate shorter-lived states, albeit with reduced counting statistics.

\section{Results}

\begin{figure*}[t!]
\begin{centering}
\includegraphics[width=18.0cm]{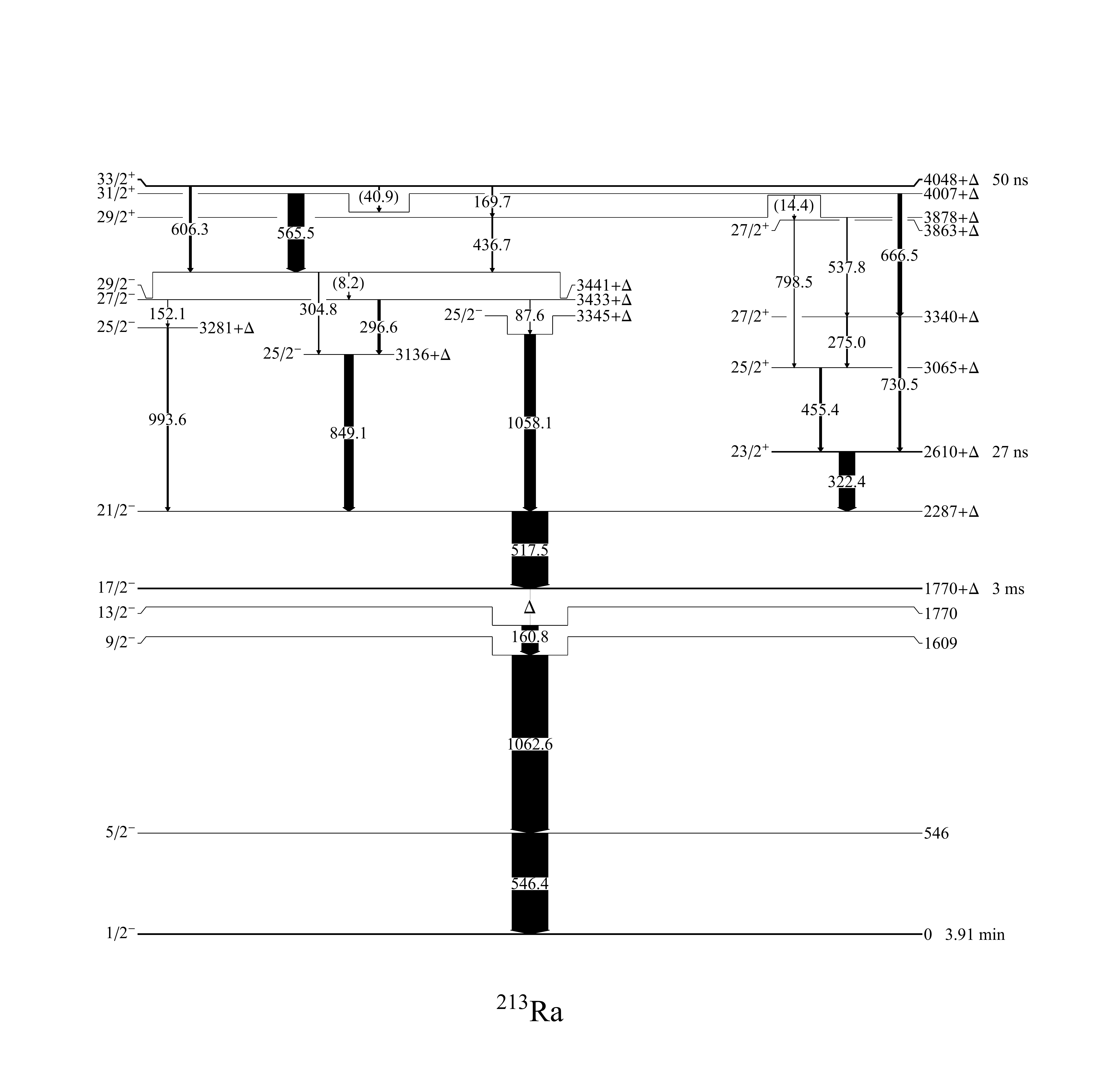}
\caption{Level scheme of $^{213}$Ra. All levels located above the $J^\pi$~=~17/2$^-$ isomer have been identified for the first time. Widths of the arrows indicate $\gamma$-ray intensities relative to the 518-keV gating transition feeding the $J^\pi$~=~17/2$^-$ isomer in the out-of-beam data. Transition widths below the $J^\pi$~=~17/2$^-$ isomer are based on the measurements in Ref.~\cite{raich-ra213}. The state lifetimes shown are mean lives.} 
\label{Ra213levelscheme}		
\end{centering}
\end{figure*}

\begin{figure*}[t!]
\includegraphics[width=18.0cm]{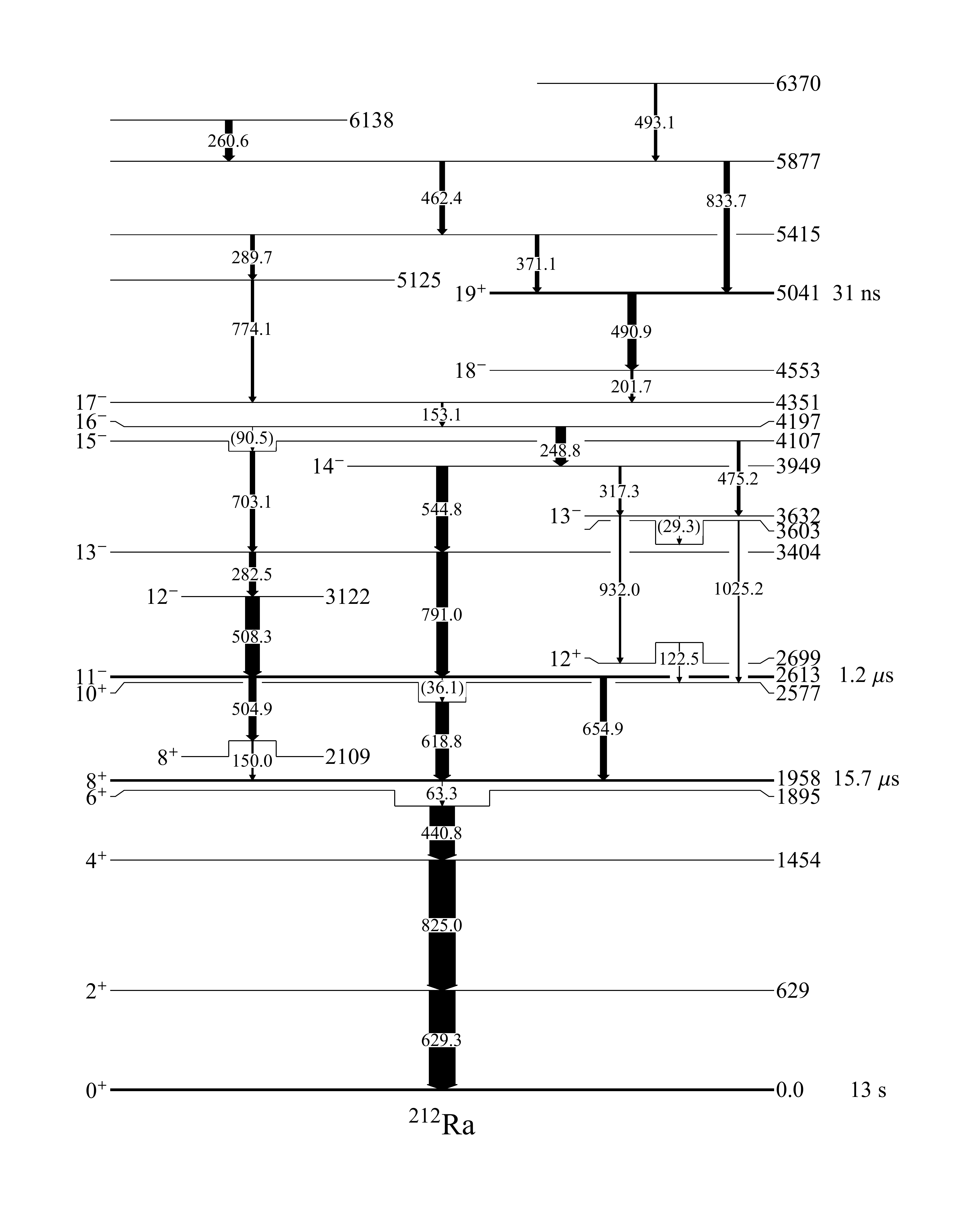}
\caption{Level scheme of $^{212}$Ra. Most of the levels above the $J^\pi$~=~11$^-$ isomer have been identified for the first time. Widths of the arrows indicate relative $\gamma$-ray intensities. Transition intensities of the cascade below the $J^\pi$~=~8$^+$ isomer were determined from a singles spectrum.  Relative intensities for transitions occurring between the $J^\pi$~=~11$^-$ and $J^\pi$~=~8$^+$ isomers were determined using a sum of gates on the 508-keV and 791-keV transitions. Above the $J^\pi$~=~$11^-$ isomer, intensities were determined using a sum of gates set on the 505-keV, 619-keV and 655-keV transitions.  State lifetimes are given as mean-lives.}
\label{Ra212levelscheme}
\end{figure*}

Evidence and justification for the proposed $^{213}$Ra and $^{212}$Ra level schemes, which extend the previous work \cite{raich-ra213,ra212}, are provided below. Inspection of $\gamma$-ray spectra recorded both `in-beam' ($-13$~ns~to~+30~ns around the beam pulse) and in various `out-of-beam' (+30~ns~to~+1700~ns after the beam pulse) gates were used to identify previously unknown, high-spin isomers in both isotopes. Characteristics of these new excited states and transitions are provided in Tables \ref{ra213 transitions}, \ref{unassignedTransitions} and \ref{ra212 transitions}. The proposed level schemes are presented in Fig.~\ref{Ra213levelscheme} and Fig.~\ref{Ra212levelscheme}. All levels located above the  $J^\pi$~=~17/2$^-$ isomer in $^{213}$Ra, and most of the levels above the  $J^\pi$~=~11$^-$ isomer in $^{212}$Ra, have been identified for the first time.

\subsection{$^{213}$Ra level scheme}

Prior knowledge \cite{raich-ra213} of the $^{213}$Ra level scheme was limited to a cascade of three $E2$ $\gamma$-ray transitions connecting the yrast  $J^\pi$~=~13/2$^-$ level to the  $J^\pi$~=~1/2$^{-}$ ground state, and evidence for an unobserved transition with energy of less than 10~keV \cite{raich-ra213} between the  $J^\pi$~=~13/2$^-$ state and the  $J^\pi$~=~17/2$^-$, $\tau$~=~3-ms isomer. 
Observation of transitions below the $J^\pi$~=~17/2$^-$ isomer confirmed the production of $^{213}$Ra in the experiment, however, the 3-ms lifetime of the isomer precluded direct correlation of any observed transitions arising from above this state with the known level structure lying below. Nevertheless, $\gamma$-ray coincidences with characteristic radium X~rays \cite{xrays}, and prior knowledge of the neighbouring radium isotopes \cite{ra214,ra215}, enabled unambiguous assignment of many new transitions to $^{213}$Ra located above the $J^{\pi}$~=~17/2$^-$ isomer.

\begin{figure*}
\includegraphics[angle=90,width=17cm]{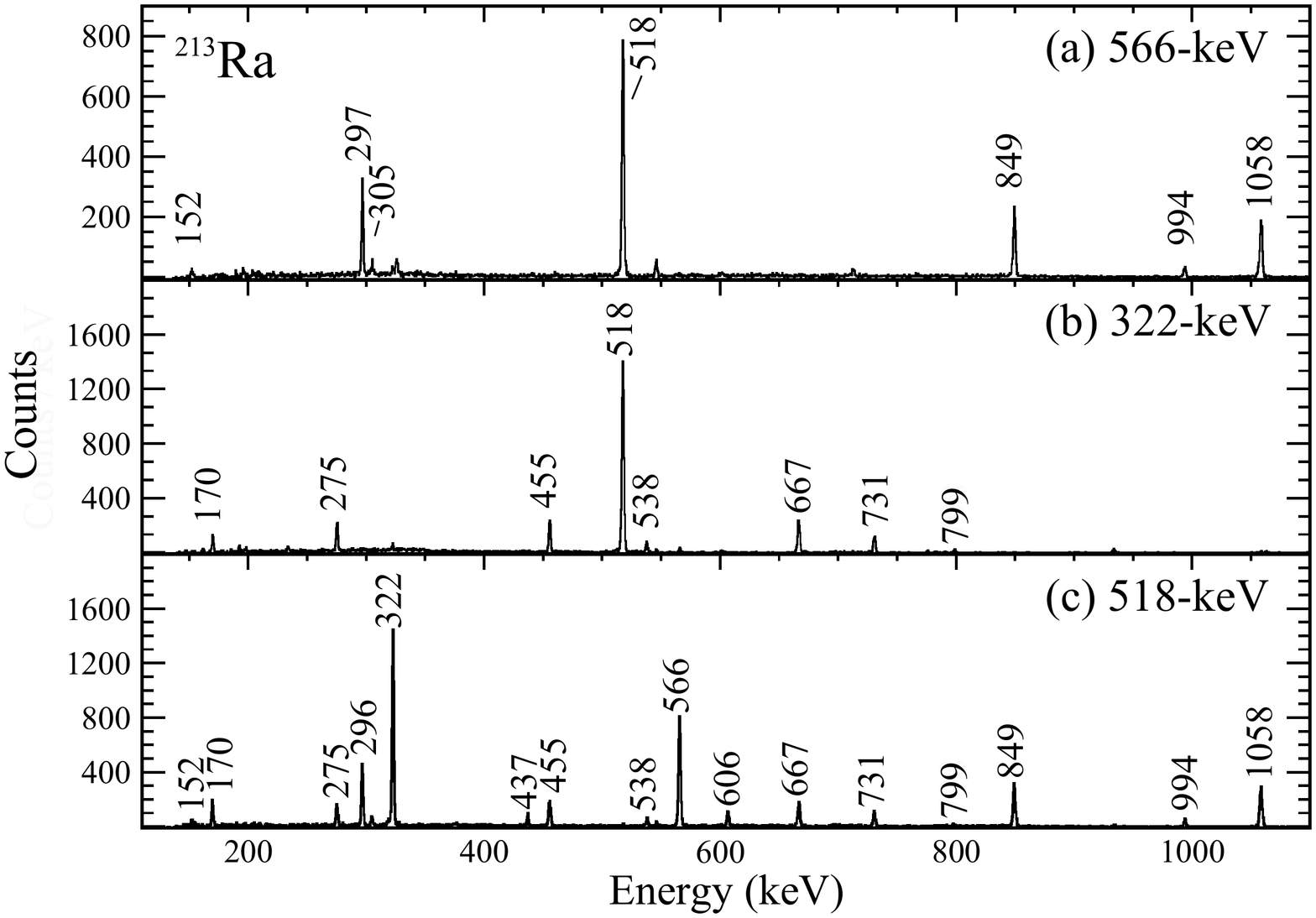}
\caption{Background-subtracted, out-of-beam (+30~ns~to~+140~ns after the beam pulse) coincidence spectra gated on the (a) 566-keV, $J^\pi$~=~$31/2^+ \rightarrow 29/2^-$, (b) 322-keV,  $J^\pi$~=~$23/2^+ \rightarrow 21/2^-$ and (c) 518-keV, $J^\pi$~=~$21/2^- \rightarrow 17/2^-$ transitions in $^{213}$Ra.}
\label{ra213Spectra}
\end{figure*}

The $\gamma$ rays observed in the in-beam data, and identified to be in coincidence with the 88.47-keV radium $K_{\alpha 1}$ X~ray \cite{xrays}, yielded a number of strong transitions that had not previously been assigned to any of the radium isotopes. The most intense of these was the 518-keV transition that, based on its intensity, has been assigned to directly feed the $J^\pi$~=~17/2$^-$ isomer. Background-subtracted, out-of-beam coincidence spectra, gated on the 566-keV, 322-keV and 518-keV transitions, are provided in Fig.~\ref{ra213Spectra}. The fact that these $\gamma$ rays are strongly populated in the out-of-beam data provides compelling evidence for the existence of further high-spin isomers in $^{213}$Ra. The coincidence spectrum generated by gating on the 518-keV $\gamma$-ray transition contains almost every new transition that is placed in the extended $^{213}$Ra level scheme.

The out-of-beam coincidence spectra, gated on the 566-keV and 322-keV transitions (Figs.~\ref{ra213Spectra}a and \ref{ra213Spectra}b), illustrate the parallel cascades that occur above the 518-keV transition. The ordering of transitions was established unambiguously from a number of crossover transitions. The exceptions are the ordering of the 994-152-keV, and 1058-88-keV cascades. Each of these could, in principle, be reversed. Their ordering was based on the in-beam intensities measured by gating on the 518-keV transition. Further evidence for their ordering was obtained via comparison with the semiempirical shell-model calculations discussed in Section \ref{Discussion}, which predict states lying close in energy to those in the proposed level scheme.

\begin{table*}[t]
\centering
\caption{New $\gamma$-ray transitions placed above the $J^\pi$~=~17/2$^-$ isomer in $^{213}$Ra. Relative intensities are normalized to that of the 518-keV $\gamma$-ray transition. Evidence for their placement in the level scheme, along with $\sigma L$ and $J^{\pi}$ assignments, are discussed in the text. The state energies contain the $+\Delta$ term due to the unknown energy of the $J^\pi$~=~$17/2^- \rightarrow 13/2^-$ transition.}
\label{ra213 transitions}
\begin{ruledtabular}
\begin{tabular}{cccccccccc}
$E_\gamma$	&	$I_\gamma$	&	$E_i - \Delta$	&	$J^\pi_i$	&	$E_f - \Delta$	&	$J^\pi_f$	&	\textit{$\sigma$}L	&	$\alpha_T$(exp.)	&	$\alpha_T$(calc.)	&	$A_2$/$A_0$	\\

\hline
(8.2)	&		&	3441 &	29/2$^-$	&	3433	&	27/2$^-$	&	$M$1\footnotemark[1]	&		&		&		\\
(14.4) 	&		&	3878	&	29/2$^+$	&	3863	&	27/2$^+$	&	$M$1\footnotemark[1]	&		&		&		\\
(40.9)     	&		&	4048	&	33/2$^+$	&	4007	&	31/2$^+$ 	&	$M$1\footnotemark[1]	&		&		&		\\
87.6(4)        	&		&	3433	&	27/2$^-$	&	3345	&	25/2$^-$	&	$M$1\footnotemark[1]	&		&		&		\\
152.1(5)	&	2.3(2)	&	3433	&	27/2$^-$	&	3281	&	25/2$^-$	&	$M$1+$E2$\footnotemark[2]	&	3.7(5)	& 3.72		&		\\
169.7(3)	&	4.4(2)	&	4048	&	33/2$^+$	&	3878	&	29/2$^+$	&	$E$2	&	1.3(2)	& 1.1		&		\\
275.0(4)	&	4.2(2)	&	3340	&	27/2$^+$	&	3065	&	25/2$^+$	&	$M$1	&	0.99(8)	&	0.9	&	$-$0.41(16) 	\\
296.6(2)	&	12.3(3)	&	3433	&	27/2$^-$	&	3136	&	25/2$^-$	&	$M$1 (+$E$2)	&	0.96(9)	&	0.74	&	$-$0.73(4) 	\\
304.8(8)	&	1.8(2)	&	3441	&	29/2$^-$	&	3136	&	25/2$^-$	&	$E$2\footnotemark[1]	&		&		&	          	\\
322.4(1)	&	44.6(6)	&	2610	&	23/2$^+$	&	2287	&	21/2$^-$	&	$E$1	&	0.11(5)	&	0.03	&	$-$0.31(5) 	\\
436.7(8)	&	2.4(2)	&	3878	&	29/2$^+$	&	3441	&	29/2$^-$	&	$E$1\footnotemark[1]	&		&		&	          	\\
455.4(4)	&	7.1(3)	&	3065	&	25/2$^+$	&	2610	&	23/2$^+$	&	$M$1 + $E$2	&		&		&	$-$1.06(6) 	\\
458.7(6)\footnotemark[3]	&		&	(4507+$\Delta^{\prime}$)	&	(37/2$^+$)	&	(4048+$\Delta^{\prime}$)	&	&	&		&		&	          	\\
517.5(1)	&	100	&	2287	&	21/2$^-$	&	1770	&	17/2$^-$	&	$E$2	&		&		&	0.26(1)  	\\
537.8(11)	&	2.0(2)	&	3878	&	29/2$^+$	&	3340	&	27/2$^+$	&	$M$1	&		&		&	$-$0.46(7)	\\
565.5(2)	&	41.0(7)	&	4007	&	31/2$^+$	&	3441	&	29/2$^-$	&	$E$1	&		&		&	$-$0.46(6)	\\
606.3(5)	&	5.4(3)	&	4048	&	33/2$^+$	&	3441	&	29/2$^-$	&	$M2$+$E3$\footnotemark[1]	&		&		&	          	\\
666.5(4)	&	10.0(4)	&	4007	&	31/2$^+$	&	3340	&	27/2$^+$	&	$E$2	&		&		&	 0.38(13) 	\\
730.5(5)	&	5.8(3)	&	3340	&	27/2$^+$	&	2610	&	23/2$^+$	&	$E$2	&		&		&	 0.38(17) 	\\
798.5(10)	&	0.7(2)	&	3863	&	27/2$^+$	&	3065	&	25/2$^+$	&	$M1$+$E2$\footnotemark[1]	&		&		&	          	\\
849.1(3)	&	24.4(6)	&	3136	&	25/2$^-$	&	2287	&	21/2$^-$	&	$E$2	&		&		&	0.38(4)	\\
993.6(7)	&	3.7(3)	&	3281	&	25/2$^-$	&	2287	&	21/2$^-$	&	$E$2	&		&		&	 0.33(14) 	\\
1058.1(2)	&	28.3(7)	&	3345	&	25/2$^-$	&	2287	&	21/2$^-$	&	$E$2	&		&		&	 0.30(5)  	\\
\end{tabular}
\end{ruledtabular}
\footnotetext[1]{Transition multipolarity implied by the determined spins and parities of initial and final states.}
\footnotetext[2]{With a mixing ratio of $\left|\delta(E2/M1)\right|=0.7(3)$. The calculated $\alpha_T$ is for $\delta=0.7$.}
\footnotetext[3]{The possible placement of this transition above an unobserved low-energy ($\Delta^{\prime}$-keV), $J^\pi$~=~$35/3^+ \rightarrow 33/2^+$ transition is discussed in the text.}
\end{table*}

\begin{table}[htb]
\centering
\caption{Transitions coincident with strong transitions in $^{213}$Ra that could not be placed in the level scheme due to insufficient statistics. These transitions do not occur following the decay of an isomeric state and bypass the isomer at 4048$+\Delta$ keV. }
\label{unassignedTransitions}
\begin{ruledtabular}
\begin{tabular}{cc}
Transition energy (keV) & Coincident transitions (keV) \footnotemark[1] \\
\hline
204& 314, 377, 397/398, 416, 447\\
314& 204, 377, 397/398, 416, 447\\
377& 204, 314, 397/398, 416, 447\\
397& 204, 314, 377, 398, 416, 447\\
398& 204, 314, 377, 397, 416, 447\\
416& 204, 377, 397/398\\
447& 204, 377, 397/398\\
459& 397/398, 565, 667
\end{tabular}
\end{ruledtabular}
\footnotetext[1]{The transitions are also coincident with the 297-keV, 518-keV, 849-keV, 994-keV and 1058-keV transitions.}
\end{table}

\begin{table*}[t]
\centering
\caption{$\gamma$-ray transitions observed above the $J^\pi$~=~8$^+$ isomer in $^{212}$Ra. Relative intensities are normalized to the 619-keV $\gamma$ ray. Evidence for placements in the level scheme, as well as $\sigma L$ and $J^{\pi}$ assignments, are discussed in the text.}
\label{ra212 transitions}
\begin{ruledtabular}
\begin{tabular}{cccccccccc}
$E_\gamma$	&	$I_\gamma$	&	$E_i$	&	$J^\pi_i$	&	$E_f$	&	$J^\pi_f$	&	\textit{$\sigma$}L	&	$\alpha_T$(exp.)	&	$\alpha_T$(calc.)	&	$A_2$/$A_0$	\\
\hline
(29.3)  	&	  	&	3632	&	13$^-$ 	&	3603	&		&		&		&		&		\\
(36.1)  	&	   	&	2613	&	11$^-$ 	&	2577	&	10$^+$	&	$E1$	&		&		&		\\
(90.5)  	&			&	4197	&	16$^-$ 	&	4107	&	15$^-$	&	$M1$	&		&		&		\\
122.5(1) 	&	  	&	2700	&	12$^+$ 	&	2577	&	10$^+$	&	$E2$	&	4.7(10)	&	4.035	&	 	\\
150.0(2) 	&	4.9(6)		&	2109	&	8$^+$  	&	1958	&	8$^+$ 	&	$M1$	&		&		&		\\
153.1(2) 	&	5.9(14) 	&	4351	&	17$^-$ 	&	4197	&	16$^-$	&	$M1$	&	5.0(3)	&	4.63	&	1.0(4)	\\
201.7(1) 	&	7.9(16)		&	4553	&	18$^-$ 	&	4351	&	17$^-$	&	$M1 + E2$\footnotemark[1]	&1.89(11)	 	&1.94		&	0.3(2) 	\\
248.8(1) 	&	35.8(40)	&	4197	&	16$^-$ 	&	3949	&	14$^-$	&	$E2$	&	0.31(4)	&	0.273	&	0.24(14)	\\
260.6(1) 	&	24.9(30) 	&	6138	&		&	5877	&	 	&		&	 	&	 	&	0.45(23) 	\\
282.5(1) 	&	23.1(31)	&	3404	&	13$^-$ 	&	3122	&	12$^-$	&	$M1$	&	0.98(5)	&	0.835	& $-$0.66(17)		\\
289.7(1) 	&	14.0(23)	&	5415	&	 	&	5125	&		&		&		&		&	 $-$0.6(3)	\\
317.3(1) 	&	7.5(18)		&	3949	&	14$^-$ 	&	3632	&	13$^-$	&	$M1$	&	0.65(13)	&	0.608	&		\\
371.1(1) 	&	13.2(24)	&	5415	&	     	&	5041	&	19$^+$	&	   	&		&		&		\\
462.4(1) 	&	17.6(33)	&	5877	&	   	&	5415	&		&	   	&		&		&	$-$0.3(4)	\\
475.2(2) 	&	10.4(24)	&	4107	&	15$^-$ 	&	3632	&	13$^-$	&	$E2$	&		&		&		\\
490.9(1) 	&	30.3(45) 	&	5041	&	19$^+$ 	&	4553	&	18$^-$	&	$E1$	&		&		&$-$0.23(18)		\\
493.1(2) 	&	9.1(22)  	&	6370	&	 	&	5877	&	 	&	   	&		&		&		\\
504.9(2) 	&	27.4(22)  &	2613	&	11$^-$ 	&	2109	&	8$^+$ 	&	$E3$	&		&		&		\\
508.3(1) 	&	51.1(58)	&	3122	&	12$^-$ 	&	2613	&	11$^-$	&	$M1 + E2$	&		&		&	$-$0.85(10)	\\
544.8(1) 	&	41.5(52) 	&	3949	&	14$^-$ 	&	3404	&	13$^-$	&	$M1$	&		&		&	$-$0.07(15)\\
618.8(1) 	&	48.9(35) 	&	2577	&	10$^+$ 	&	1958	&	8$^+$ 	&	$E2$	&		&		&		\\
654.9(2) 	&	23.7(23)	&	2613	&	11$^-$ 	&	1958	&	8$^+$	&	$E3$	&		&		&		\\
703.1(1) 	&	16.5(39)	&	4107	&	15$^-$ 	&	3404	&	13$^-$	&	$E2$	&		&		&0.7(2)		\\
774.1(1) 	&	8.9(26)		&	5125	&	 	&	4351	&	17$^-$	&	   	&		&		&		\\
791.0(1) 	&	40.1(59)	&	3404	&	13$^-$ 	&	2613	&	11$^-$	&	$E2$	&		&		&0.48(14)	\\
833.7(1) 	&	19.8(44)  	&	5877	&	 	&	5041	&	19$^+$	&	   	&		&		&		\\
932.0(1) 	&	5.1(23)  	&	3632	&	13$^-$ 	&	2700	&	12$^+$	&	$E1$	&		&		&		\\
1025.2(2)	&	3.6(30)	&	3603	&		&	2577	&	10$^+$	&	   	&		&		&		\\
\end{tabular}
\end{ruledtabular}
\footnotetext[1]{With a mixing ratio of $\left|\delta(E2/M1)\right|=0.43^{+12}_{-13}$. The calculated $\alpha_T$ is for $\delta=0.43$.}
\end{table*}

\begin{figure*}
\includegraphics[angle=-90,width=17cm]{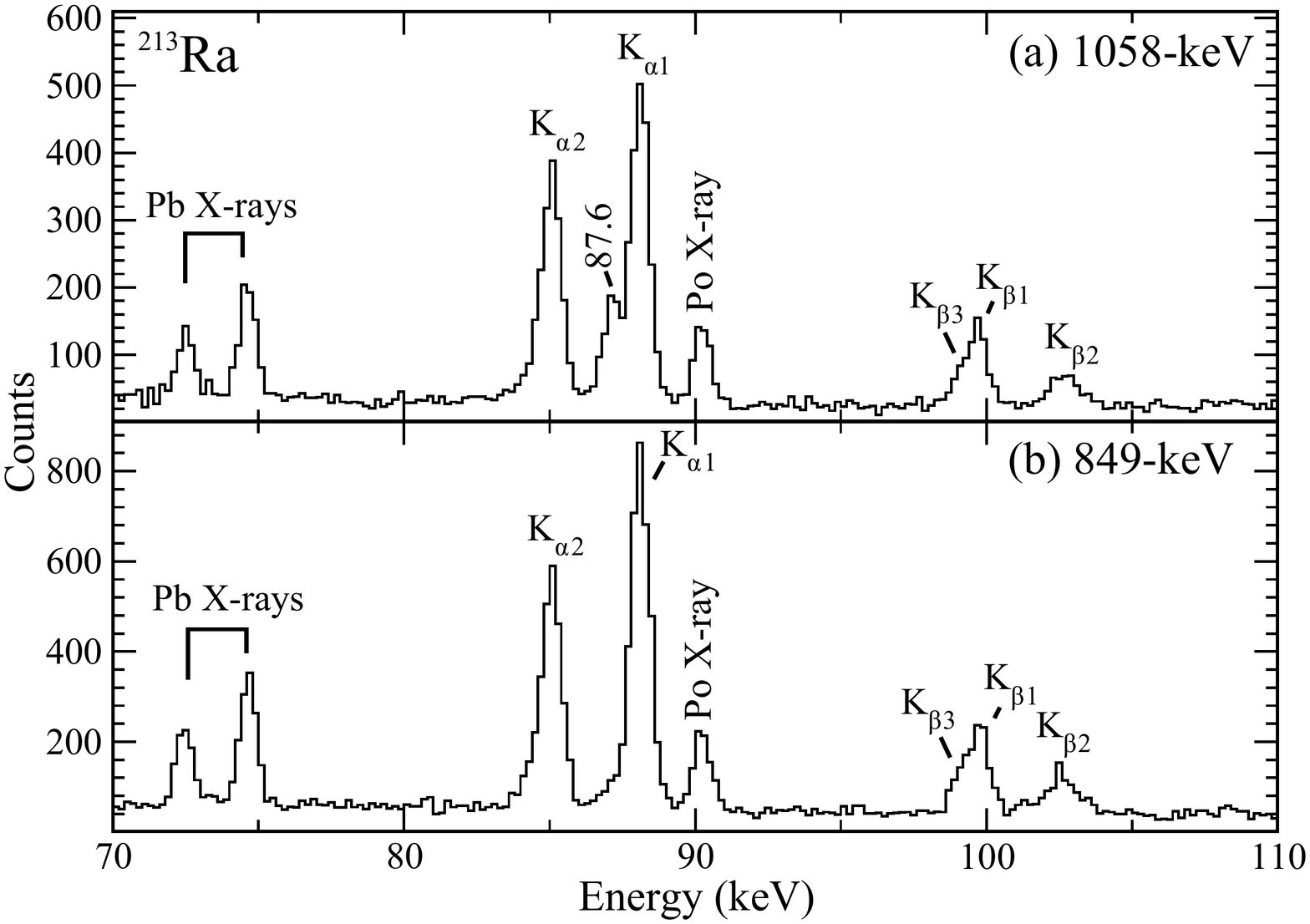}
\caption{Background-subtracted, LEPS coincidence spectra for $^{213}$Ra, gated on the (a) 1058-keV, $J^\pi$~=~$25/2^+ \rightarrow 21/2^-$, (b) 849-keV, $J^\pi$~=~$25/2^+ \rightarrow 21/2^-$ transitions in $^{213}$Ra observed in the HPGe detectors. Lines labelled with K designations refer to the radium X-rays, other X-rays from random coincidences are also labeled.}
\label{LEPS}
\end{figure*}

The 88-keV transition was observed directly, and resolved from the radium X-rays, by projecting the $\gamma$ rays detected in the LEPS detectors that were coincident with the 566-keV and 1058-keV transitions detected in the HPGe detectors; it was not observed in gates on the parallel 297-849-keV cascade. LEPS spectra with gates on the 1058-keV and 849-keV transitions are shown in Fig.~\ref{LEPS}. The meeting point of the two significant parallel cascades (one passing through exclusively positive parity states, the other through predominantly negative parity states) was established by gating on the 518-keV transition. The 437-keV and 538-keV $\gamma$ rays present in this gate were not in coincidence with the strong 566-keV transition that feeds the $J^\pi$~=~29/2$^-$ state. The 437-keV $\gamma$ ray is coincident with both members of the 305--849-keV cascade that feeds the $J^\pi$~=~21/2$^-$ level. The 538-keV transition was found to be in coincidence with the 731-keV and 322-keV transitions, which feed and depopulate the $J^\pi$~=~23/2$^+$ isomer. The energy sum of the 437--305--849-keV cascade is equal to the sum of the 538--731--322-keV cascade within uncertainties, which, when combined with the $\gamma$-$\gamma$ coincidence relationships, supports the placement of a level at 3878+$\Delta$~keV. The energy sums of the 566--305--849-keV cascade and the 667--731--322-keV cascade are equal within uncertainties, placing a new level linking the two parallel cascades at 4007+$\Delta$~keV. The 8.2-keV and 14.4-keV transitions are unobserved gaps inferred from energy sums and comparisons between coincidence spectra.

The energy sum of the 170-keV and 437-keV transitions is equal to the energy of the 606-keV transition within uncertainties, suggesting these are parallel cascades de-exciting an additional state located at 4048+$\Delta$ keV. This state was found to be isomeric and the measurement of its lifetime is discussed below. An unobserved 41-keV transition is placed between the 4048+$\Delta$-keV and 4007+$\Delta$-keV levels that feeds towards the 566-keV and 667-keV decay sequences that were discussed above. Projecting `early' $\gamma$~rays detected 30 ns to 150 ns prior to the 322-keV, 518-keV and 566-keV transitions showed an additional, strong 459-keV $\gamma$ ray that is evidently not emitted following the decay of an isomeric state. This transition is attributed to a level located above the $J^\pi$~=~33/2$^+$ isomer, possibly at 4507+$\Delta$~keV. We have not shown the 459-keV line on the level scheme because the semiempirical shell-model calculations strongly suggest that there may be an additional unobserved low-energy transition from a $J^\pi$~=~35/2$^+$ state immediately above the $J^\pi$~=~33/2$^+$ isomer, which could not be either confirmed or excluded. In addition to the 459-keV transition, several transitions (listed in Table \ref{unassignedTransitions}) were observed in-beam in a decay path bypassing the isomer at 4048+$\Delta$ keV. Due to insufficient statistics these transitions could not be confidently placed in the $^{213}$Ra level scheme; however, coincidences with strong $^{213}$Ra $\gamma$ rays confirm their assignment to this nucleus.

\subsection{$^{212}$Ra level scheme}

\subsubsection{\textbf{Transitions below the 4351-keV $J^\pi$~=~17$^-$ state}}

\begin{figure*}
\begin{center}
\includegraphics[angle=-90,width=17cm]{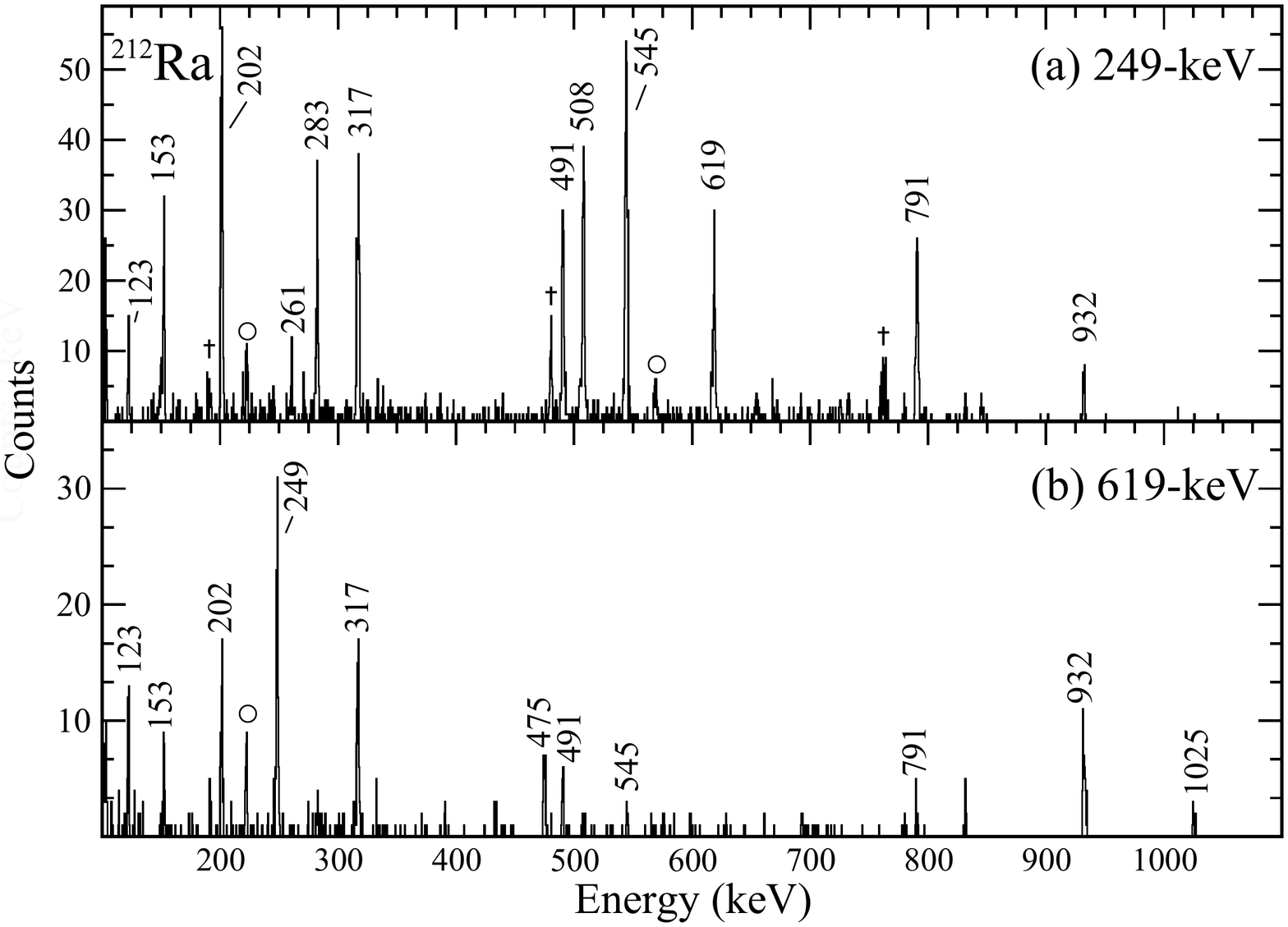}
\caption{Background-subtracted, out-of-beam (+30~ns~to~+140~ns after the beam pulse) coincidence spectra gated on the (a) 249-keV,  $J^\pi$~=~$16^- \rightarrow 14^-$ and (b) 619-keV,  $J^\pi$~=~$10^+ \rightarrow 8^+$ transitions in $^{212}$Ra. Contaminant $\gamma$-ray transitions indicated by the symbol $\dagger$ arise due to the reaction product $^{209}$Rn. The transitions labeled with $\circ$ are coincident with a number of $^{212}$Ra $\gamma$~rays and radium x rays but could not be placed in the level scheme.}
\label{ra212spectra-1}
\end{center}
\end{figure*}

\begin{figure*}
\begin{center}
\includegraphics[angle=-90,width=17cm]{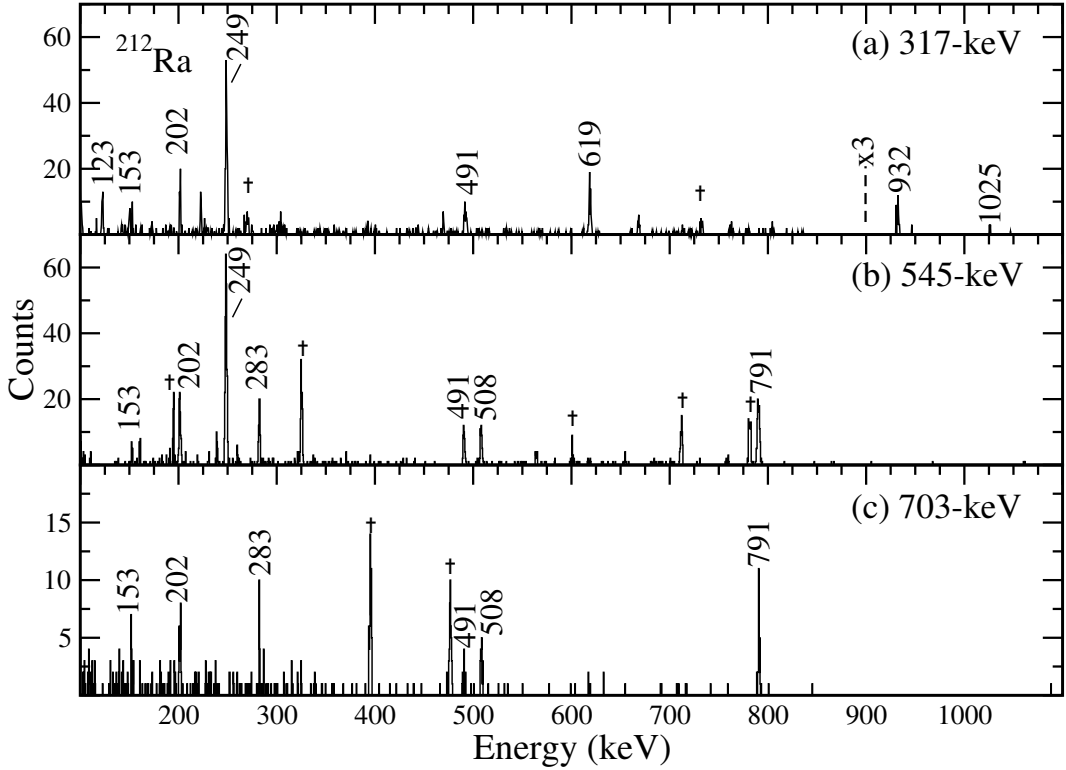}
\caption{Background-subtracted, out-of-beam (+30~ns~to~+140~ns after the beam pulse) coincidence spectra gated on the (a) 317-keV, $J^{\pi}$~=~$14^- \rightarrow 13^-$, (b) 545-keV, $J^{\pi}$~=~$14^- \rightarrow 13^-$, and (c) 703-keV, $J^{\pi}$~=~$15^- \rightarrow 13^-$ transitions in $^{212}$Ra. Contaminant $\gamma$-ray transitions, indicated by the symbol $\dagger$, arise primarily due to the reaction products $^{206}$Po and $^{209}$Po.}
\label{ra212spectra-2}
\end{center}
\end{figure*}

Background-subtracted, coincidence spectra gated on the 249-keV and 619-keV transitions are provided in Fig.~\ref{ra212spectra-1}. Despite lying below the $J^\pi$~=~11$^-$ isomer, gating on the 619-keV $\gamma$ ray isolates transitions that follow the decay of higher-spin isomers populated in the reaction due to the 123-keV decay branch bypassing the isomer. Spectroscopic evidence for new parallel decay cascades located below the state at 4351 keV was obtained by gating on the 317-keV, 545-keV and 703-keV $\gamma$-ray transitions in the out-of-beam data, as shown in Fig.~\ref{ra212spectra-2}. The 153-keV $\gamma$~ray is common to all three spectra, placing a meeting point of these three cascades below this transition. The 249-keV transition is common to the 317-keV and 545-keV gates, identifying a meeting point between these two branches. The placement of a 91-keV gap between the 4198-keV and 4107-keV levels is inferred from the observed coincidence between 703-keV and 153-keV transitions. The energy sums of the 123--932--475--91-keV, 36--791--703--91-keV and 36--791--545--249-keV cascades are equal within the experimental uncertainties, supporting the placement of the unobserved 91-keV gap. The order of transitions can be established unambiguously throughout this section of the level scheme due to the number of intersecting cascades. The only exception is the 283--508-keV cascade, the order of which has been based upon the relative intensities of the two transitions when projecting $\gamma$-rays early with respect to the 619-keV $\gamma$~ray. The gate on the 317-keV transition in Fig.~\ref{ra212spectra-2}a shows a loss of intensity in the 317--932--123-keV and 317--29--1025-keV cascades relative to the 249-keV and 619-keV $\gamma$-ray transitions, above and below these sequences. The existence of additional decay paths out of the 3632-keV state is thus implied but, due to poor statistics, no candidate transitions could be identified in the $\gamma$-ray spectra.

\subsubsection{\textbf{Transitions above the 4351-keV $J^\pi$~=~17$^-$ state}}

\begin{figure*}
\begin{center}
\includegraphics[width=17cm]{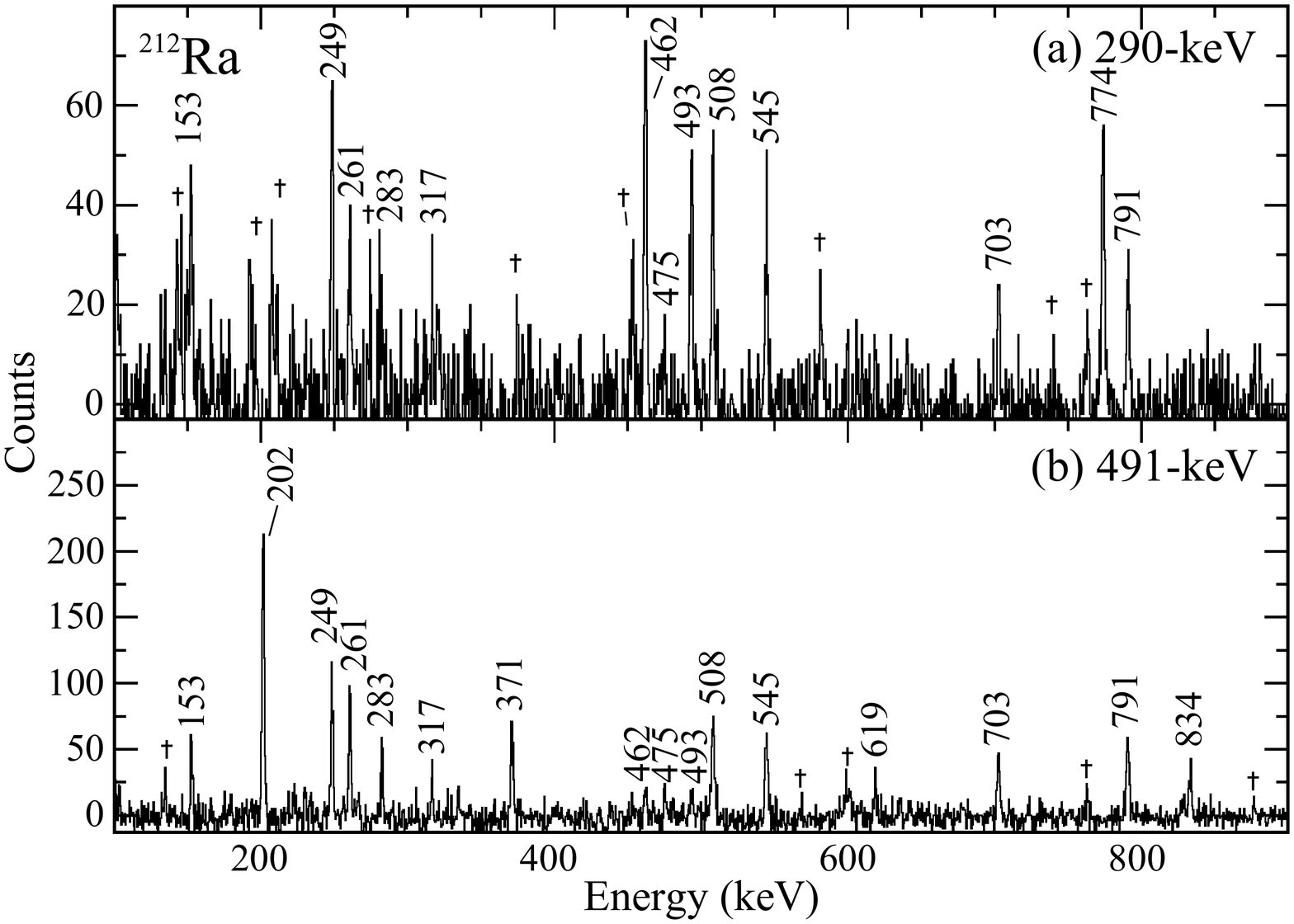}
\caption{Background-subtracted, in-beam ($-13$~ns~to~+30~ns around the beam pulse) coincidence spectra gated on the (a) 290-keV and (b) 491-keV, $J^{\pi}$~=~$19^+ \rightarrow 18^-$ transitions in $^{212}$Ra that demonstrate the presence of short-lived states above the level at 5041 keV. Contaminant $\gamma$-ray transitions indicated by the symbol $\dagger$ come from a number of excited nuclei populated in the reaction, for example $^{204}$Pb.}
\label{ra212spectra-3}
\end{center}
\end{figure*}

An additional nine $\gamma$-ray transitions connecting seven excited states above the 4351-keV, $J^\pi$~=~17$^-$ state have been placed in the $^{212}$Ra level scheme. In-beam coincidence spectra, gated on the 491-keV and 290-keV transitions, are shown in Fig.~\ref{ra212spectra-3}. These serve as evidence of the parallel cascades located above the 4351-keV level. The 491-keV $\gamma$~ray is present in both the in-beam and ``short" out-of-beam data while the transitions located above it are only identified in the in-beam data, which indicates that the 5041-keV state is isomeric. Measurement of the mean life of this isomer is discussed below. The placement of excited states located above the 5041-keV isomer is unambiguous; however, the ordering of the 290--774-keV cascade through the proposed 5125-keV state could be reversed.

\subsection{Conversion Coefficients}

\begin{figure}[b!]
\centering
\includegraphics[width=\columnwidth, clip]{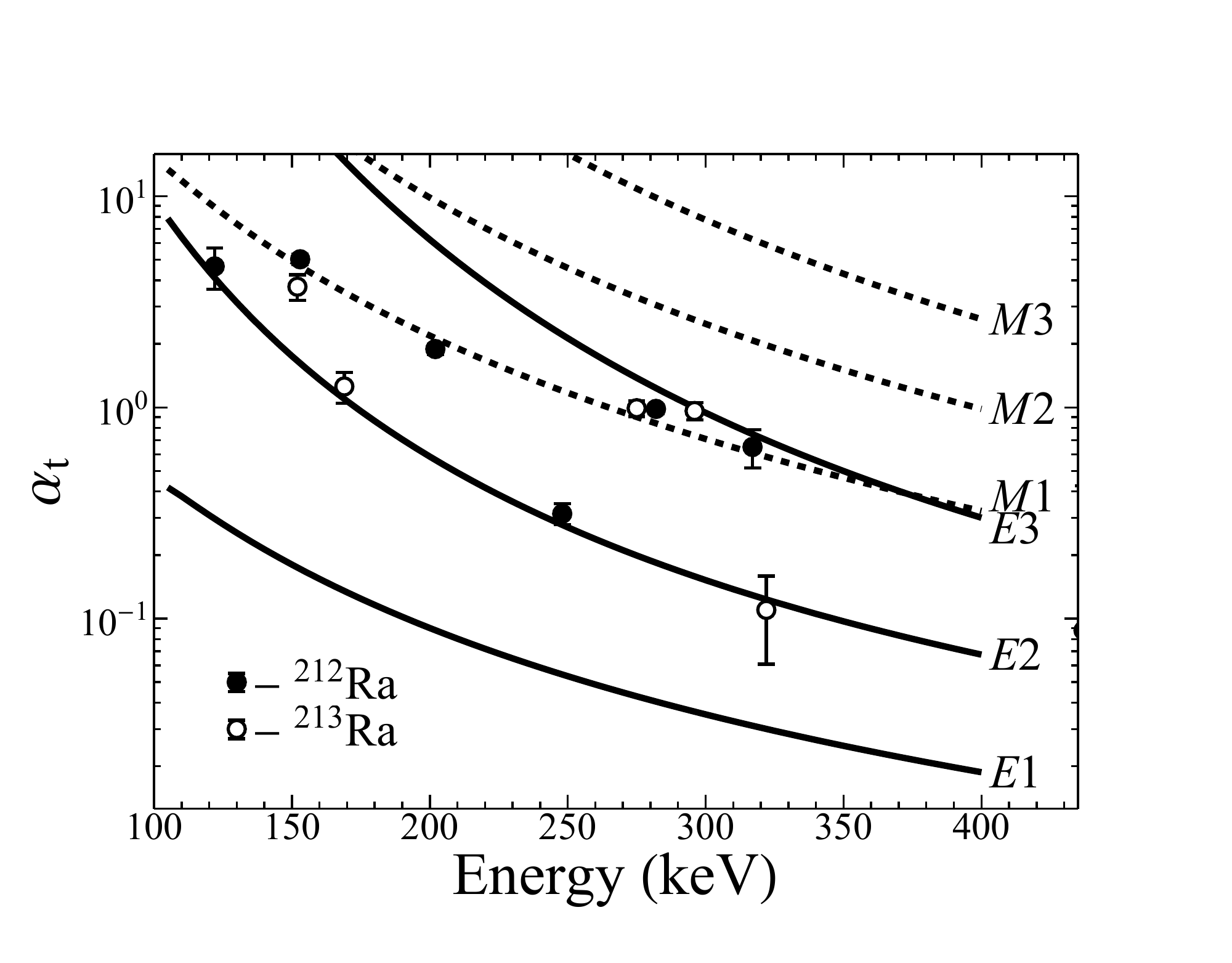}
\caption{Total internal conversion coefficients deduced from intensity balances of low-energy transitions in $^{212}$Ra (filled circles) and $^{213}$Ra (open circles). Solid and dashed lines are theoretical values for electric and magnetic transitions, respectively, calculated using BRICC \cite{bricc}.}
\label{ICC_plot}
\end{figure}

Total internal conversion coefficients were deduced from the intensity balance across levels. For instance, in the lower panel of  Fig.~\ref{ra213Spectra}, the measured yield of the 152-keV transition that feeds the 3281+$\Delta$-keV, $J^\pi$~=~25/2$^-$ state, would be expected to balance that of the subsequent 994-keV transition that depopulates it. Internal conversion coefficients were extracted for 12 transitions in $^{212}$Ra and $^{213}$Ra from similar considerations. Comparison of the experimental values to theoretical calculations using the BRICC code \cite{bricc} constrained the multipolarity of the transitions, as illustrated in Fig.~\ref{ICC_plot}. In $^{213}$Ra, most conversion coefficients were determined by gating on the 518-keV transition out-of-beam, which gives clean spectra following the decay of the 4048+$\Delta$-keV, $J^\pi$~=~33/2$^+$ isomer. In $^{212}$Ra, conversion coefficients were determined by gating on a variety of transitions due to the large number of pathways identified throughout the level scheme.

\subsection{Angular anisotropies}

\begin{figure}[t!]
\centering
\includegraphics[width=\columnwidth, clip]{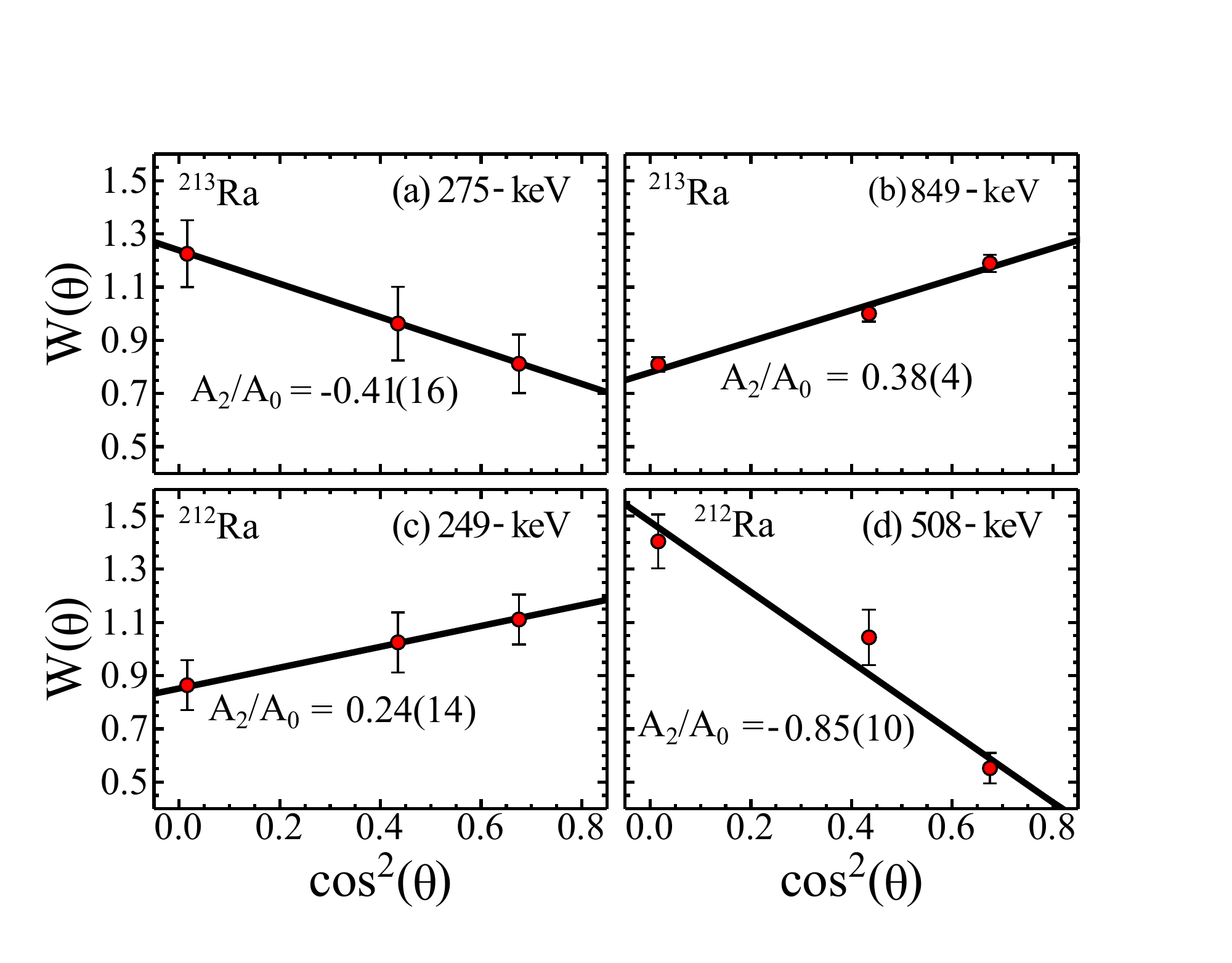}
\caption{Measured angular distributions plotted for the (a) 275-keV and (b) 849-keV transitions in $^{213}$Ra, and the (c) 249-keV and (d) 508-keV (d) transitions in $^{212}$Ra. The data points correspond to the three detector pairs positioned in the vertical plane relative to the beam axis at $\pm$34$\degree$, $\pm$48$\degree$ and $\pm$82$\degree$.}
\label{ADplot}
\end{figure}

\begin{figure}[h]
\centering
\includegraphics[width=\columnwidth, clip]{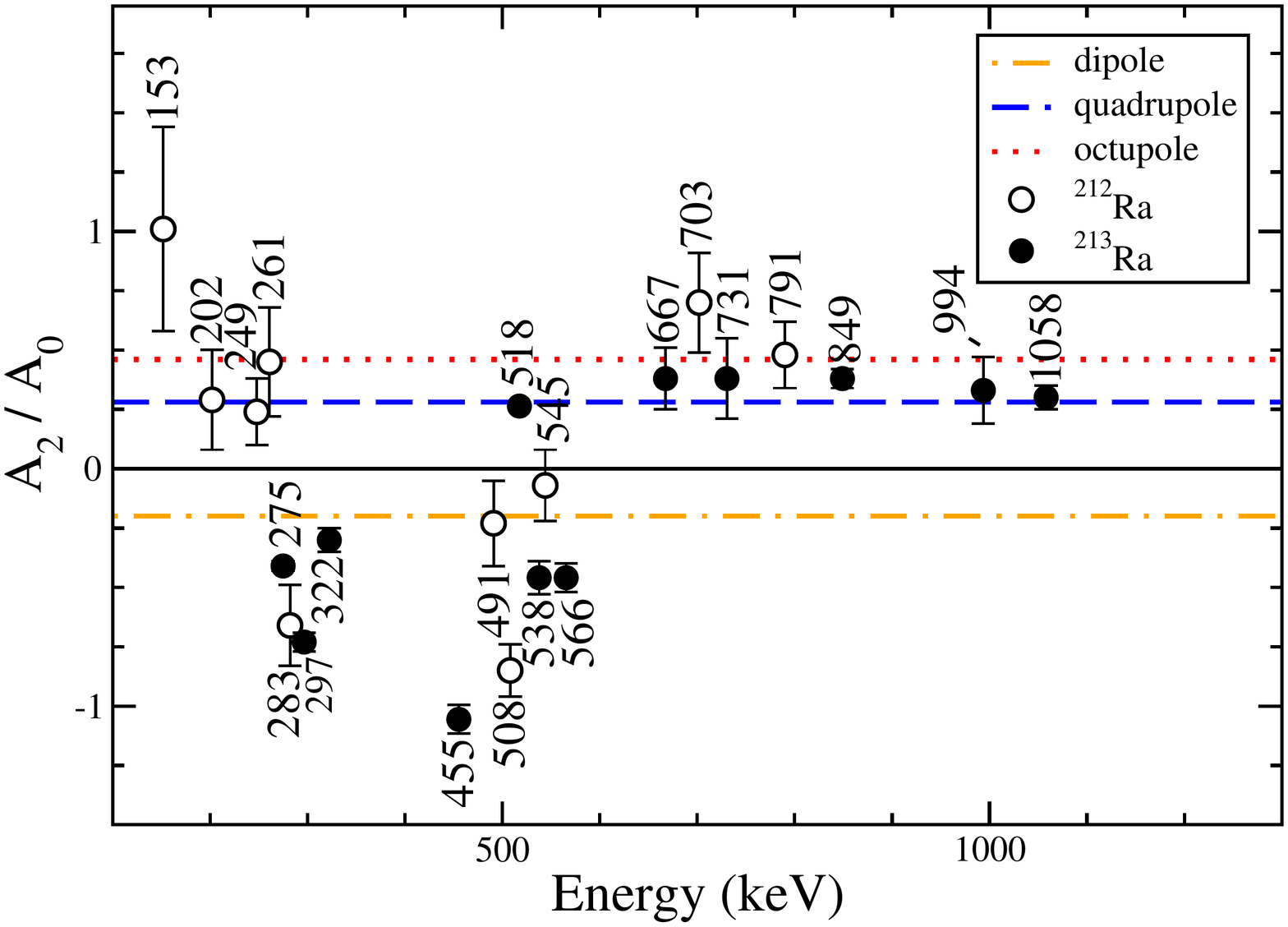}
\caption{Extracted values of $A_2 / A_0$ obtained for $\gamma$-ray transitions in $^{212}$Ra (open circles) and $^{213}$Ra (closed circles). Dashed horizontal lines at $-0.2$, 0.26 and 0.46 are the calculated values for pure dipole, quadrupole and octupole transitions assuming an alignment due to a Gaussian $m$-substate distribution with $\sigma/J$ = 0.3 (see, for example, Ref.~\cite{MorinagaP54}). A significant deviation from these lines is an indication of a mixed multipolarity transition.}
\label{A2plot}
\end{figure}

Measured angular anisotropies provided information on the angular-momentum change due to transitions in $^{212}$Ra and $^{213}$Ra. The $A_2/A_0$ coefficients extracted from fitting $W(\theta)$ = $A_0$ $+$ $A_2 P_2(\cos\theta)$ to the data are listed in Tables \ref{ra213 transitions} and \ref{ra212 transitions}. Selected fits to the experimental data are displayed in Fig.~\ref{ADplot} and extracted $A_2/A_0$ coefficients are displayed in Fig.~\ref{A2plot}. 
Further physical considerations can facilitate the determination of the electric or magnetic nature of many identified transitions. There is a well-established occurrence of $E3$ transitions from isomeric states in neighboring nuclides \cite{Vincent-fr210, ra214, ra215}. Consequently, any transitions found to exhibit a large $A_2/A_0$ coefficient without a discernible state lifetime can be considered to be $E2$ in nature. As will be discussed in more detail below, low-energy $E1$ transitions in these nuclei are normally associated with the decay of isomeric levels. Thus, in most cases, $A_2/A_0$ values indicating a dipole character without a measurable lifetime can be considered to be $M1$ transitions. Significant deviation of the measured $A_2/A_0$ coefficients from the expected values can indicate mixed-multipolarity transitions. However, with only three detector angles available and, therefore, no measurement of the corresponding $A_4$ term, the exact mixing ratio cannot be determined. The 297-keV and 455-keV transitions in $^{213}$Ra, and the 508-keV transition in $^{212}$Ra (see Figure \ref{ADplot}d) appear to have mixed $E2$/$M1$ multipolarity.

\subsection{Lifetime measurements}

\begin{figure}[t!]
\includegraphics[width=\columnwidth]{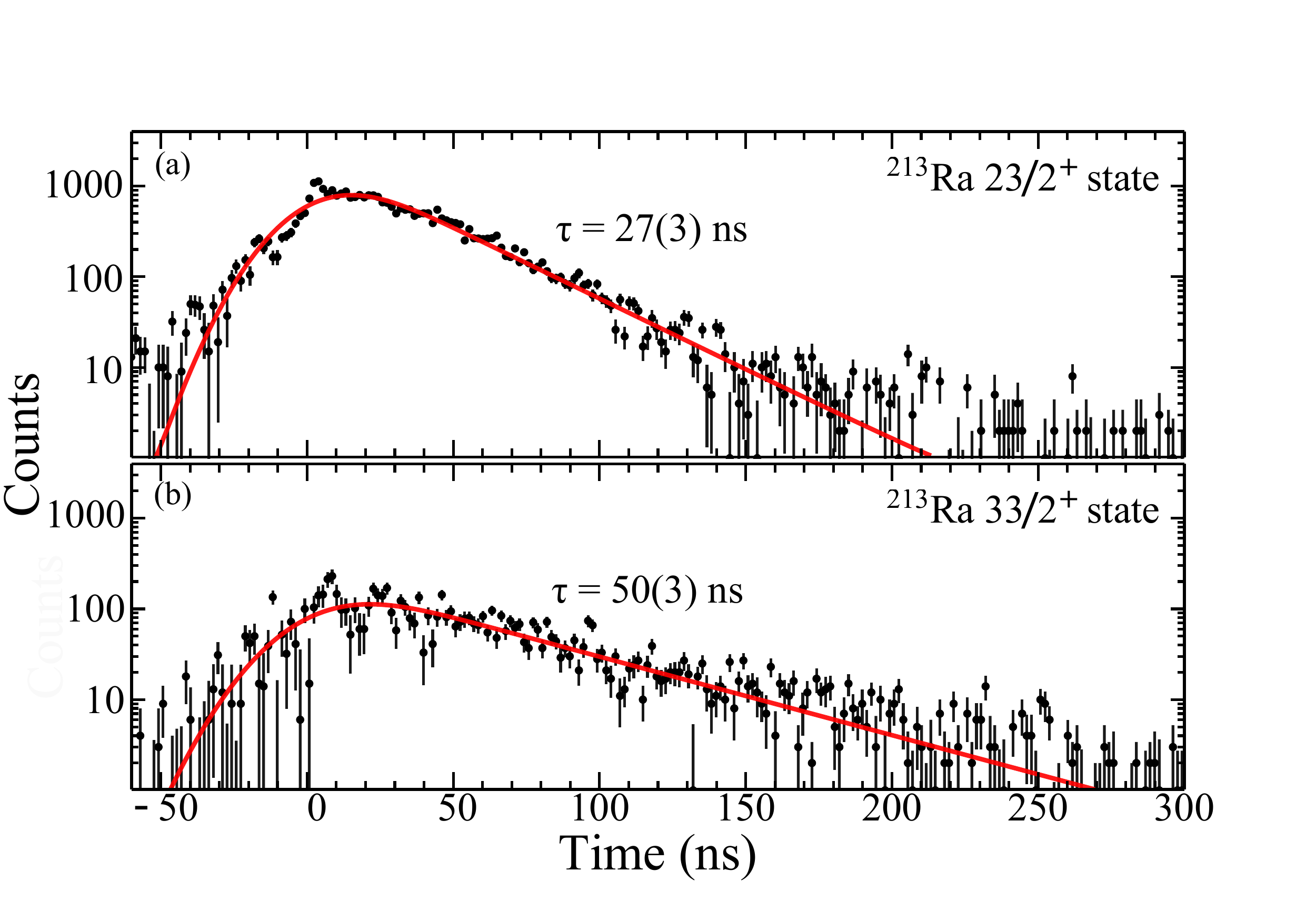}
\caption{The $\gamma$-$\gamma$ time-difference spectra showing the mean lives of the (a) $J^\pi$~=~23/2$^+$ and (b) $J^\pi$~=~33/2$^+$ isomers in $^{213}$Ra (see text for gating transition details).}
\label{ra213lifetimecurves}
\end{figure}

The lifetime of the $4048+\Delta$-keV isomer in $^{213}$Ra, determined from the time difference between the 459-keV transition feeding the isomer and a sum of gates on the 170-keV, 566-keV and 606-keV transitions below the isomer, is 50(3) ns, as shown in Fig.~\ref{ra213lifetimecurves}. Lifetime measurements with respect to the beam pulse for the 566-keV and 667-keV transitions showed no discernible difference, demonstrating that this choice of gating transitions does not bias the lifetime curve. A similar approach was used to measure the lifetime of the 2610+$\Delta$-keV state. The time difference between the 455-keV and 731-keV feeding transitions and the 322-keV decaying transition yields a lifetime of 27(3) ns.

\begin{figure}[h]
\centering
\includegraphics[width=\columnwidth]{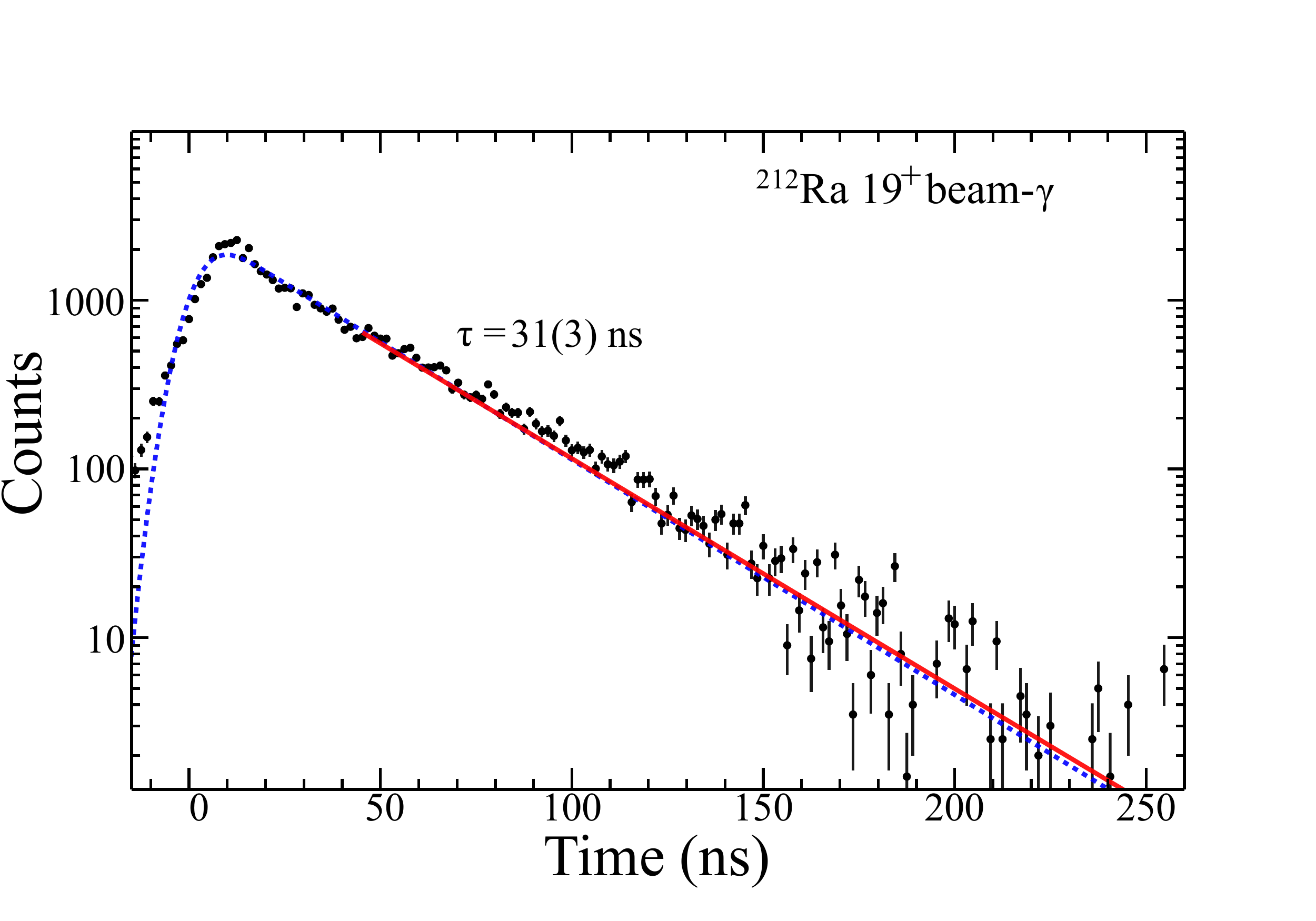}
\caption{Histogram of the number of 491-keV $\gamma$ rays versus time between beam pulses, which define $t = 0$. A mean life of 31(3)~ns was obtained by combining results from fitting the decay curve with either the slope (solid red), or including the full prompt and slope components (dotted blue).}
\label{ra212lifetimecurves}
\end{figure}

The 31(3) ns lifetime of the 5041-keV state in $^{212}$Ra, shown in Fig.~\ref{ra212lifetimecurves}, has been measured with reference to the beam pulse by gating on the 491-keV transition and projecting the resulting time spectrum. An additional time difference measurement was attempted, however, even a sum of many time difference spectra had insufficient statistics. Measurements of the $J^\pi$~=~11$^-$ state lifetime confirmed the literature value of $\tau = 1.2(2)$~$\mu$s \cite{ra212}, but the maximum beam pulse separation of 1.7 $\mu$s precluded obtaining an independent result of comparable precision.

\subsection{Spin-parity assignments}

Spin and parity assignments are known from prior studies up to the $J^\pi$~=~17/2$^-$ isomer in $^{213}$Ra \cite{neyens-ra213}, and to the $J^\pi$~=~13$^-$ state in $^{212}$Ra \cite{ra212}. Spins and parities have been assigned to the majority of new states identified in the present work based on a combination of measured internal conversion coefficients, angular anisotropies, and state lifetimes, along with constraints required by crossover transitions.

\subsubsection{\textbf{Spin-parity assignments in $^{213}$Ra}}

The 2287+$\Delta$-keV state is assigned as $J^\pi$~=~21/2$^-$ on the basis that the 518-keV transition connecting this state to the $J^\pi$~=~17/2$^-$ isomer was determined to be an $E2$ transition from the measured $A_2/A_0$ value and lack of a measurable state lifetime. The $J^\pi$~=~21/2$^-$ state is fed by four $\gamma$ rays with energies of 322~keV, 849~keV, 994~keV and 1058~keV. The three higher energy transitions are assigned as stretched $E2$ on the basis of their $A_2/A_0$ values and the absence of measurable lifetimes for the states that they depopulate. Thus the states at 3136+$\Delta$~keV, 3281+$\Delta$~keV and 3345+$\Delta$~keV are all assigned $J^{\pi}$~=~25/2$^-$. The internal conversion coefficient of the 152-keV transition supports an $M1+E2$ assignment with a mixing ratio of $\left|\delta(E2/M1)\right|=0.7(3)$. The 297-keV transition's $A_2/A_0$ value and its conversion coefficient indicate $M1$ multipolarity, while the lack of an observable lifetime associated with the 8-keV gap between the $3433+\Delta$-keV and $3441+\Delta$-keV states also implies $M1$ multipolarity. Thus all three $J^{\pi}$~=~25/2$^-$ states are fed by a $J^{\pi}$~=~27/2$^-$ state at $3433+\Delta$ keV, with a $J^{\pi}$~=~29/2$^-$ state 8 keV higher at $3441+\Delta$ keV.
The 8--152--994-keV, 8--297--849-keV, and 8--88--1058-keV cascades all converge at the $J^{\pi}$~=~29/2$^-$, 3441+$\Delta$-keV state. The 566-keV transition feeding the $J^{\pi}$~=~29/2$^-$ state has a negative $A_2/A_0$ value, suggesting a dipole transition. There is no measurable lifetime associated with the 4007+$\Delta$-keV state, which is assigned $J^{\pi} = 31/2^+$ from the spin and parity assignments made in the parallel branch depopulating this state. Hence the 566-keV transition must have $E1$ multipolarity, i.e. $31/2^+ \rightarrow 29/2^-$.

The 27(3)-ns lifetime of the 2610+$\Delta$-keV level and the measured negative $A_2/A_0$ value of the 322-keV transition provide compelling evidence that it is an $E1$ transition, with the measured value of the conversion coefficient, $\alpha_T=0.11(5)$, only 1.6$\sigma$ from the expected $E1$ value of 0.03 and very far from the expected $M1$ value of 0.584. Thus, $J^{\pi}$~=~23/2$^+$ is assigned to the 2610+$\Delta$-keV level. The 3065+$\Delta$-keV level is assigned $J^{\pi}$~=~25/2$^+$ on the basis of the $M1$ character of the 455-keV $\gamma$ ray. The internal conversion coefficient for the 275-keV transition supports $M1$ nature and the angular distribution of the 731-keV transition supports a quadrupole assignment. Therefore, the 3340+$\Delta$-keV state is assigned $J^\pi$~=~27/2$^+$. The 538-keV and 667-keV transitions have been assigned as $M1$ and $E2$ on similar arguments so that the 3878+$\Delta$-keV state is assigned as $J^\pi$~=~29/2$^+$ and the 4007-keV state is assigned as $J^\pi$~=~31/2$^+$. The known spins and parities below the $J^\pi$~=~29/2$^+$ level imply the 437-keV, 799-keV and 14-keV transitions are $E1$, $M1$ and $M1$, respectively.

The 170-keV transition depopulating the isomer at 4048+$\Delta$-keV has a measured total conversion coefficient consistent with that of an $E2$ transition, which implies $J^\pi$ = 33/2$^+$ for this isomer. Hence, the 41-keV transition is $M1$ and the 606-keV transition is $M2$, or more likely mixed $M2+E3$ (see below). As discussed below, the $M1$ nature of the 41-keV transition is reinforced by the relative fraction of strength carried by this $\gamma$ ray compared to the 606-keV $M2+E3$ or the 170-keV $E2$ transitions.

\subsubsection{\textbf{Spin-parity assignments in $^{212}$Ra}}

Assigning spins and parities to excited states in $^{212}$Ra  was more challenging than in $^{213}$Ra due to fission competition and hence reduced statistics, particularly for levels above 4197~keV. No spin or parity assignments could be made for levels located above the 5041-keV isomer. The present work confirms the spin assignments made to ten excited states in $^{212}$Ra by Kohno \textit{et al.}~\cite{ra212}, adding parity assignments to two of these levels, although there are significant differences between their level scheme and the present one for states above the $J^\pi$~=~13$^-$, 3404-keV level. 

The internal conversion coefficient inferred from intensity balances for the 123-keV transition suggests it is an $E2$ transition, thus connecting a $J^{\pi}$~=~12$^+$ level at 2700 keV to the 2577-keV, $J^{\pi}$~=~10$^+$ level. The $A_2/A_0$ value of the 508-keV transition is consistent with $M1$ nature and so the previously reported \cite{ra212} 3122-keV, $J$~=~12 state is confirmed and assigned a negative parity. The spin and parity of $J^{\pi}$~=~13$^-$ for the 3404-keV level is confirmed by the measured internal conversion coefficient for the $M1$ 283-keV transition connecting it to the 3122-keV, $J^{\pi}$~=~12$^-$ level. The measured $A_2/A_0$ value for the intense 791-keV transition depopulating this $J^\pi$~=~$13^-$ state and feeding the 2613-keV, $J^{\pi}$~=~11$^-$ isomer is consistent with an $E2$ multipolarity.

The $A_2/A_0$ value of the 545-keV transition is consistent with $M1$ nature and so the 3949-keV state is assigned $J^{\pi}$~=~14$^-$. The internal conversion coefficient of the 249-keV transition feeding this state suggests it is an $E2$ transition and so the 4197-keV state is assigned $J^{\pi}$~=~16$^-$. From its measured internal conversion coefficient, the 317-keV transition that depopulates the 14$^-$ level was found to be an $M1$ transition and so a $J^{\pi}$ = 13$^-$ assignment is made to the 3632-keV state. The 3632-keV, $J^{\pi}$~=~13$^-$ level is connected to the 2700-keV, $J^{\pi}$~=~12$^+$ level by a 932-keV $\gamma$-ray transition. An $E1$ assignment is required for this transition to be consistent with the spin-parity assignments discussed above.

Measured internal conversion coefficients for the 153-keV and 202-keV transitions support $M1$ assignments to both and so the 4351-keV and 4553-keV levels are assigned $J^{\pi}$~=~17$^-$ and $J^{\pi}$~=~18$^-$, respectively. Although the $A_2/A_0$ values measured for the 153-keV and 202-keV transitions are inconsistent with pure $M1$ assignments, introducing a mixing ratio of $\left|\delta(E2/M1)\right| = 0.43^{+12}_{-13}$ reproduces the angular distribution measured for the 202-keV transition. With this mixing ratio, the theoretical internal conversion coefficient becomes 1.94, bringing it closer to the measured value. The same technique could not be applied to the 153-keV transition, which is hampered by statistical uncertainties; the unusually large $A_2/A_0$ value in this case has been disregarded and the spin-parity assignment based upon the internal conversion coefficient alone.

The 491-keV transition that feeds the 4553-keV level has a negative $A_2/A_0$ value, which, when considered in conjunction with the measured state lifetime of 31(3)~ns, implies $E1$ multipolarity. Therefore, a $J^{\pi}$~= ~19$^+$ assignment has been made to the isomeric state at 5041 keV.

\subsection{Transition strengths}

Transition strengths determined from the isomeric-state lifetimes measured for $^{213}$Ra and $^{212}$Ra are shown in Table \ref{transitionstrengths}, along with the strengths for some related transitions in neighbouring nuclei \cite{ra214,rn210,rn211,rn212,rn211a}.

As expected, since $E1$ transitions are not possible between the valence orbitals in the major shells, the $E1$ transition strengths are all very weak, typically of order $10^{-7}$~W.u. In $^{212}$Ra, the $E3$ decays of the 2613-keV $J^{\pi}$~=~11$^-$ state to the two lower $J^{\pi}$~=~8$^+$ states are observed. In $^{213}$Ra, the $J^{\pi}$~=~33/2$^+$ isomer is depopulated by retarded $M1$ and $E2$ transitions, as well as a transition that is likely of mixed $M2+E3$ multipolarity. These transition strengths will be discussed in further detail below.

\begin{table*}[t]
\centering
\caption{Transition strengths for isomeric decays in $^{212}$Ra and $^{213}$Ra. Transition strengths in $^{214}$Ra as well as $^{210,211,212}$Rn have been included for comparison \cite{ra214,rn210,rn211,rn211a,rn212}.}
\label{transitionstrengths}
\begin{ruledtabular}
\begin{tabular}{cccccccc}
Nucleus &	$E_{\rm{level}}$ & $J_i$ &$\tau$	& $E_\gamma$ & $\sigma L$ & Transition Strength & Reference\\
        &	(keV)						 &			 &  	(ns)	  & (keV) 		 &					  & (W.u.) \\
\hline\\[-0.3cm]
$^{212}$Ra	&	2613		&11$^-$			&	$1.2(2) \times 10^3$\footnotemark[1]	&505 	&$E3$	&14.4(33)	& present work			\\
			&								&			&					&655 	&$E3$	&2.1(6)		& present work	  	\\
			&								&			&					&36	 	&$E1$	&9.6(2)$\times$10$^{-7}$ & present work\\
			&	5041 &19$^+$	&31(3)                 &491 	&$E1$	&7.4(7)$\times$10$^{-8}$ & present work\\
\hline\\[-0.3cm]
$^{213}$Ra	&	2610+$\Delta$ & 23/2$^+$	&	27(3)		&323 	&$E1$	&2.7(3)$\times$10$^{-7}$ & present work	\\
			&	4048+$\Delta$	&33/2$^+$					&50(3)		&170 	&$E2$	&5.3(6)$\times$10$^{-3}$ & present work	\\
			&								&		&			&41	 	&$M1$	&1.76(11)$\times$10$^{-4}$ & present work	\\
			&								&		&			&606 	&$M2$\footnotemark[2]	&2.5(2)$\times$10$^{-4}$ & present work	\\
			&								&		&			&606 	&$E3$\footnotemark[2]	&36(3) & present work	\\
\hline\\[-0.3cm]
$^{214}$Ra&2683	&11$^-$	&426(10)						&609		&$E3$	&21.7(6) & \cite{ra214}	\\
			&					&	&									&818 		&$E3$	&3.1(1)	 & \cite{ra214}			\\
			&4810			&18$^+$	&1.1(3)						&409		&$E1$	&1.2(3)$\times$10$^{-6}$	& \cite{ra214}	\\
			&					&	&									&663		&$E1$	&4(1)$\times$10$^{-7}$ & \cite{ra214}		\\
			&6530			&(24$^+$)	&2.3(4) 		  	&48			&$M1$	&$\sim 4\times$10$^{-3}$\footnotemark[3] & \cite{ra214}	\\
			&					&	&									&240		&$E2$	&$\sim 0.2$\footnotemark[3] & \cite{ra214}	\\
\hline\\[-0.3cm]
$^{210}$Rn	&2563 &11$^-$ &92(5) 		&186	&$E1$	&3.5(2)$\times$10$^{-7}$ & \cite{rn210}	\\
						&			&				&						&898	&$E3$	&2.7(3) & \cite{rn210}\\
\hline\\[-0.3cm]
$^{211}$Rn	&2650+$\Delta$&23/2$^+$	&9.6(4) 		&503	&$E1$	&2.2(1)$\times$10$^{-7}$ & \cite{rn211,rn211a}	\\
						&		&	&		        &1073   &$E3$   &1.1(4) & \cite{rn211}\\
\hline\\[-0.3cm]
$^{212}$Rn	&2761 &11$^-$&8(3) 		&106	&$E1$	&2.1(1)$\times$10$^{-5}$	 & \cite{rn212} \\
\end{tabular}
\end{ruledtabular}
\footnotetext[1]{Lifetime from Ref.~\cite{ra212}.}
\footnotetext[2]{Alternative multipolarities. Strengths assume either a pure $M2$ or a pure $E3$ transition.}
\footnotetext[3]{Tentative values since the initial state was not firmly assigned in Ref.~\cite{ra214}. }
\end{table*}

\section{Discussion}
\label{Discussion}

Due to their proximity to the $N = 126$ neutron shell closure, it is expected that single-particle excitations will be dominant in $^{213}$Ra and $^{212}$Ra. There are limited large basis calculations available for high-spin states in nuclei beyond $^{208}$Pb \cite{LSSM.PhysRevC.67.054310}. However, there has been considerable success with a semiempirical approach to several isotopes of Rn, Fr, and Ra near $N=126$ \cite{Po208,Po209,po210,Po212,Bayer1999,Rn209.POLETTI1985,rn210,rn211,rn211a,Rn211-12,rn212,Rn213,Rn214,Vincent-fr210,Fr211-12-13,Fr214,ra214,ra215}.
The experimental level schemes of $^{212}$Ra and $^{213}$Ra were therefore compared with semiempirical shell model calculations. We designate the calculations as ``semiempirical" because they are based on empirical single-particle energies and empirical two-body interactions, as far as possible. An additional approximation is also made: configuration mixing is excluded. Thus the calculations include a diagonalization over alternative angular momentum couplings for a given orbital occupation (configuration), but do not allow the nucleons to change their orbits.
The Hamiltonian for this model can be written as usual as
\begin{equation}
\label{Hshell}
H \ = \ \sum_{i} H_{0}(i) + \sum_{i<j} V(ij) ,
\end{equation}
where $H_0$ is the single-particle contribution and $V$ the two-body residual interaction. Introducing the approximation that the residual interaction does not change or mix configurations, the excitation energy (relative to the ground state of $^{208}$Pb) can be written as
\begin{equation}
\label{Eshell}
E \ = \ \sum_{i} \left\langle i|H_{0}|i\right\rangle + \sum_{i<j}\left\langle ij|V|ij\right\rangle,
\end{equation}
where $\langle H_{i} \rangle = \langle i|H_{0}|i\rangle$ represents the empirical single-particle energies and $\langle V_{ij} \rangle = \langle ij|V|ij\rangle$  the empirical two-body residual interactions. 
Experimental values for $\langle H_{i}\rangle$ and $\langle V_{ij} \rangle$ are tabulated in Ref.~\cite{Bayer1999} and references therein. Where empirical interactions are not available they were taken from Ref.~\cite{KuoandHerling}.
We discuss first the case of $^{213}$Ra, which has a single neutron hole relative to the $N=126$ shell closure, and second consider $^{212}$Ra, which has two neutron holes.

\subsection{Shell model calculations and structure of $^{213}$Ra}

\subsubsection{Weak-coupling approximation.}

\begin{figure*}[]
\begin{center}
\includegraphics[width=17cm]{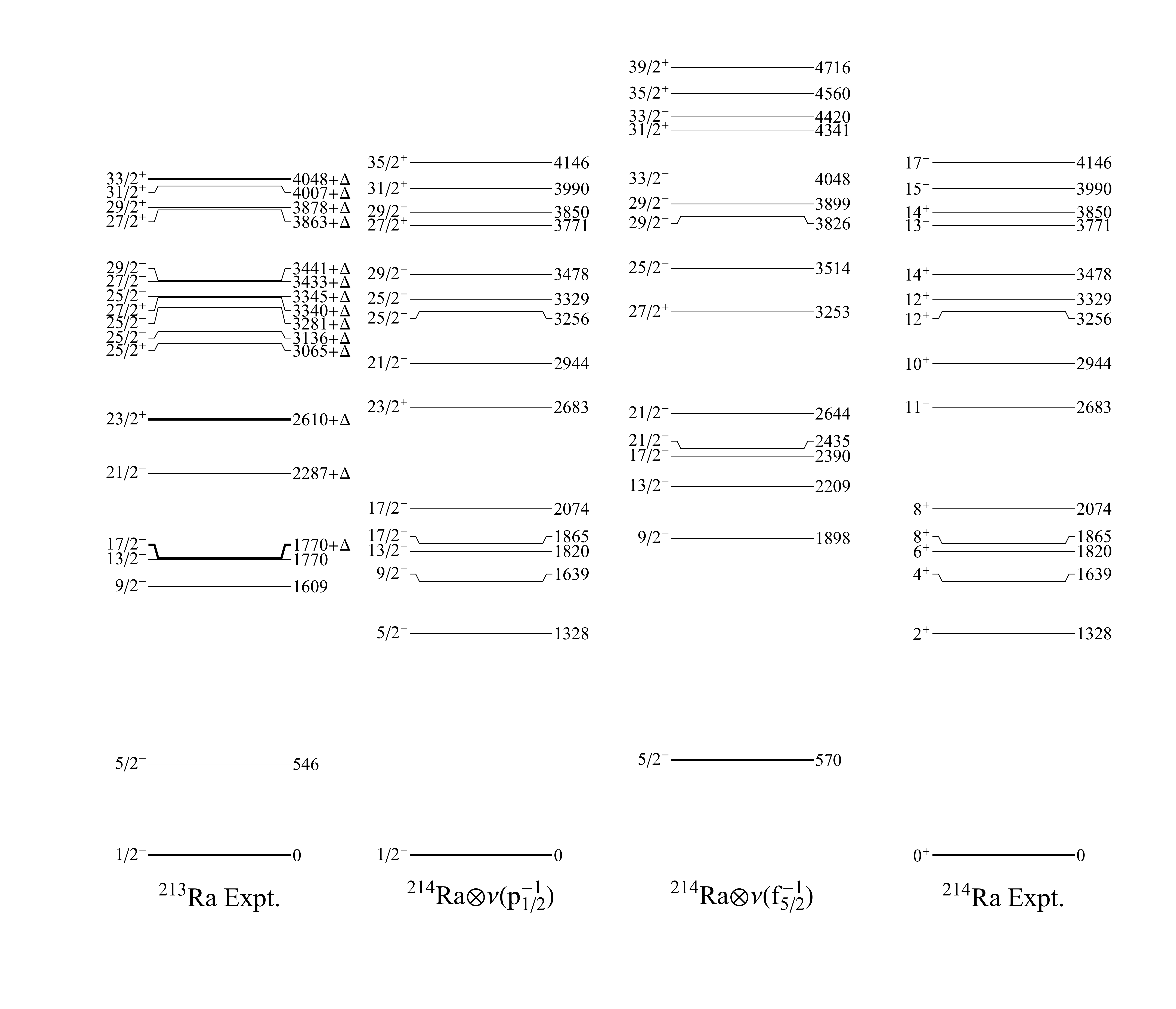}
\caption{Calculated level schemes for the weak coupling of either a $\nu p_{1/2}^{-1}$ or $\nu f_{5/2}^{-1}$ neutron hole to the known excited states in $^{214}$Ra \cite{ra214}, which is shown to the right for reference. The $^{213}$Ra level scheme is shown to the left. Isomeric states in $^{213}$Ra are indicated by thicker lines.}
\label{simpleShell}
\end{center}
\end{figure*}

The observed states in $^{213}$Ra can be formed by the coupling of a $\nu p_{1/2}$ or $\nu f_{5/2}$ neutron hole to excited states in $^{214}$Ra. Before performing the detailed semiempirical shell model calculations, it is instructive to begin with a simplified calculation of the level scheme of $^{213}$Ra by weakly coupling a neutron hole in either the $\nu p_{1/2}$ or $\nu f_{5/2}$ orbit to the observed level structure in $^{214}$Ra. We assume that (1) the resultant state in $^{213}$Ra has the maximum spin coupling, and (2) the residual interactions are equal in all states (as a consequence of the weak-coupling approximation). Thus the excited states associated with
$\nu p_{1/2}$ have the same excitation energies as in $^{214}$Ra but have the spin increased by $\frac{1}{2} \hbar$. Similarly, the states associated with the $\nu f_{5/2}$ hole have the $^{214}$Ra level spins increased by $\frac{5}{2} \hbar$ and their energies shifted up by 570 keV (the excitation energy of the first-excited $J^{\pi}$~=~$5/2^-$ state in $^{207}$Pb). Figure \ref{simpleShell} shows the results of this calculation.

As evident from Figure~\ref{simpleShell}, many of the excited states observed in $^{213}$Ra exhibit clear relationships to the $^{214}$Ra level scheme. Focusing first on states associated with levels in $^{214}$Ra up to the first $J^{\pi}$~=~8$^+$ state, nominally from the $\pi h_{9/2}^6$ configuration, we see that the $J^{\pi}$~=~1/2$^-$ ground state is associated with $(^{214}{\rm Ra}; J_p=0^+_1) \otimes \nu p_{1/2}^{-1}$.  ($J_p$ represents the spin of the protons.) Likewise, the sequence of negative parity states with  $\Delta J=2$ from $J^{\pi}$~=~9/2$^-$ to $J^{\pi}$~=~17/2$^-$ is associated with coupling the $\nu p_{1/2}$ hole to the $J^{\pi}$~=~4$^+$, $J^{\pi}$~=~6$^+$ and $J^{\pi}$~=~8$^+$ members of the nominal $\pi h_{9/2}^6$ configuration in $^{214}$Ra. The observed $J^{\pi}$~=~5/2$^-$ state is associated with $(^{214}{\rm Ra}; J_p=0^+_1) \otimes \nu f_{5/2}^{-1}$ , whereas the alternative $(^{214}{\rm Ra}; J_p=2^+_1) \otimes \nu p_{1/2}^{-1}$ is not observed because it is non-yrast. The $J^{\pi}$~=~21/2$^-$ level can be assigned as $(^{214}{\rm Ra}; J_p=8^+) \otimes \nu f_{5/2}^{-1}$. There is a second $J^{\pi}$~=~8$^+$ state in $^{214}$Ra which has no analog in the observed $^{213}$Ra level scheme.

The $J^{\pi}$~=~23/2$^+$ level at 2610 + $\Delta$ keV can be associated with the $(^{214}{\rm Ra}; J_p=11^-) \otimes \nu p_{1/2}^{-1}$ configuration, where the $^{214}$Ra $J^{\pi}$~=~11$^-$ level is nominally from the configuration $\pi h_{9/2}^5 i_{13/2}$. However, above this level, the identification of corresponding experimental and calculated levels becomes more challenging. In the case of the yrast $J^{\pi}$~=~25/2$^+$ state, there is no predicted level; presumably this level originates from a  $(^{214}{\rm Ra}; J_p=12^-) \otimes \nu p_{1/2}^{-1}$ coupling, but, as discussed further below, the relevant experimental level in $^{214}$Ra is not observed. In other cases there are several possible calculated configurations for the observed states, and configuration mixing must be expected. For example, the $J^{\pi}$~=~9/2$^-$ states from $(^{214}{\rm Ra}; J_p=4^+) \otimes \nu p_{1/2}^{-1}$ and $(^{214}{\rm Ra}; J_p=2^+)\otimes \nu f_{5/2}^{-1}$, are quite close in energy.

Three $J^{\pi}$~=~25/2$^-$ states are observed near 3.2 MeV excitation energy. Two of these can be explained by the coupling of a $\nu p_{1/2}^{-1}$ neutron hole to the two $J^{\pi}$~=~12$^+$ states in $^{214}$Ra. The third arises from the coupling of a $\nu f_{5/2}^{-1}$ neutron hole to the $J^{\pi}$~=~10$^+$ state and exists, as in experiment, on the order of 100~keV higher than the other two. The $J^{\pi}$~=~27/2$^-$ state in $^{213}$Ra has no corresponding state in $^{214}$Ra, although it is likely to arise from a $\nu p_{1/2}^{-1}$ neutron hole coupled to a $(h_{9/2}^5\otimes f_{7/2})_{13^+}$ proton configuration. The $J^{\pi}$~=~29/2$^-$ state near 3.4 MeV could be produced by four possible couplings: a $\nu p_{1/2}^{-1}$ neutron hole coupled to one of two $J^{\pi}$~=~14$^+$ states; or the $\nu f_{5/2}^{-1}$ neutron hole coupled to one of the two $J^{\pi}$~=~12$^+$ states. The lowest in energy of these possible configurations, $(^{214}{\rm Ra}; J_p=14^+) \otimes \nu p_{1/2}$, nominally $\pi (h_{9/2}^5 f_{7/2})_{14^+} \otimes \nu p_{1/2}$, is closest in energy to the observed state, and is therefore the most likely candidate.

Two $J^{\pi}$~=~33/2$^-$ states are predicted that have not been observed in $^{213}$Ra. However, weak decay paths were observed which bypass the $J^{\pi}$~=~33/2$^+$ isomer through states that could not be placed in the experimental level scheme (see Table \ref{unassignedTransitions}). Some of these unplaced states could be candidates for the unobserved predicted states.

The splitting of the experimental level scheme of $^{213}$Ra into separate cascades connecting either positive or negative states is evident in Fig.~\ref{Ra213levelscheme}, between the $J^{\pi}$~=~21/2$^-$ and $J^{\pi}$~=~33/2$^+$ states. The weak-coupling calculations better reproduce the negative parity sequence than the positive parity states, where predictions are missing for some observed states.

Overall, of the seven positive-parity states observed in $^{213}$Ra, five are predicted by the weak-coupling calculation. As already noted, the $J^{\pi}$~=~23/2$^+$ isomer in $^{213}$Ra is explained by the coupling of the $\nu p_{1/2}^{-1}$ neutron to the $J^{\pi}$~=~11$^-$ state in $^{214}$Ra, i.e. it has nominal configuration $\pi (h_{9/2}^5\otimes i_{13/2})_{11^-} \otimes \nu p_{1/2}^{-1}$. Above this state, both $J^{\pi}$~=~27/2$^+$ states are reproduced, one from $\pi (h_{9/2}^5\otimes i_{13/2})_{13^-} \otimes \nu p_{1/2}^{-1}$ and the other from $\pi (h_{9/2}^5\otimes i_{13/2})_{11^-} \otimes \nu f_{5/2}^{-1} $. The energy spacing between the predicted $J^{\pi}$~=~27/2$^+$ states is consistent with the experimental difference of $\sim 520$~keV. However, predictions for the $J^{\pi}$~=~25/2$^+$, $J^{\pi}$~=~29/2$^+$ and $J^{\pi}$~=~33/2$^+$ states in $^{213}$Ra are missing. Clearly, to predict these positive parity states in $^{213}$Ra, the simple weak-coupling model requires the observation of the associated negative-parity state in $^{214}$Ra. The relevant states have $J^{\pi}$~=~10$^-$, $J^{\pi}$~=~12$^-$, $J^{\pi}$~=~14$^-$ and $J^{\pi}$~=~16$^-$. These states are not observed in $^{214}$Ra because the observed near-yrast cascade from $J^{\pi}$~=~17$^-$ to $J^{\pi}$~=~11$^-$ in $^{214}$Ra by-passes the even-$J$, odd-parity states.

At higher excitation energies, two $J^{\pi}$~=~31/2$^+$ and two $J^{\pi}$~=~35/2$^+$ states are predicted, but only one of each is observed. The $\nu p_{1/2}^{-1}$ coupling to $J^{\pi}$~=~15$^-$ well reproduces the $J^{\pi}$~=~31/2$^+$ state, and, as the other level is higher and non-yrast, its non-observation is to be expected. In contrast, the short-comings of this simple approach are becoming apparent in the predictions of the $J^{\pi}$~=~35/2$^+$ states. The $\nu f_{5/2}^{-1}$ neutron-hole coupling to $J^{\pi}$~=~15$^-$ closely matches the experimental energy of the experimental $J^{\pi}$~=~35/2$^+$ state, but the predicted lower state is not observed, which is problematic because if it were yrast, as predicted, it should have been observed.

To sum up this section, the weak-coupling model calculations have allowed us to assign likely configurations to most of the observed excited states in $^{213}$Ra. The good agreement achieved for states up to the $J^{\pi}$~=~23/2$^+$ level  is particularly important to confirm that the level structure observed above the $J^{\pi}$~=~17/2$^-$ isomer is indeed in $^{213}$Ra, since we are unable to perform $\gamma$-$\gamma$ coincidences across this long-lived isomer. One limitation of the model is that it relies on the observation of all parent states in $^{214}$Ra, a condition that is not always met. Another limitation is that it assumes all residual proton-neutron interactions are equal, which leads to poor predictions of level energies in some cases.

\subsubsection{Semiempirical shell-model calculations}

More comprehensive semiempirical shell-model calculations were performed. These calculations provide the means to assign configurations to the experimentally observed excited states and to discuss the transition strengths in cases where they have been measured.

\begin{figure*}[]
\begin{center}
\includegraphics[width=17cm]{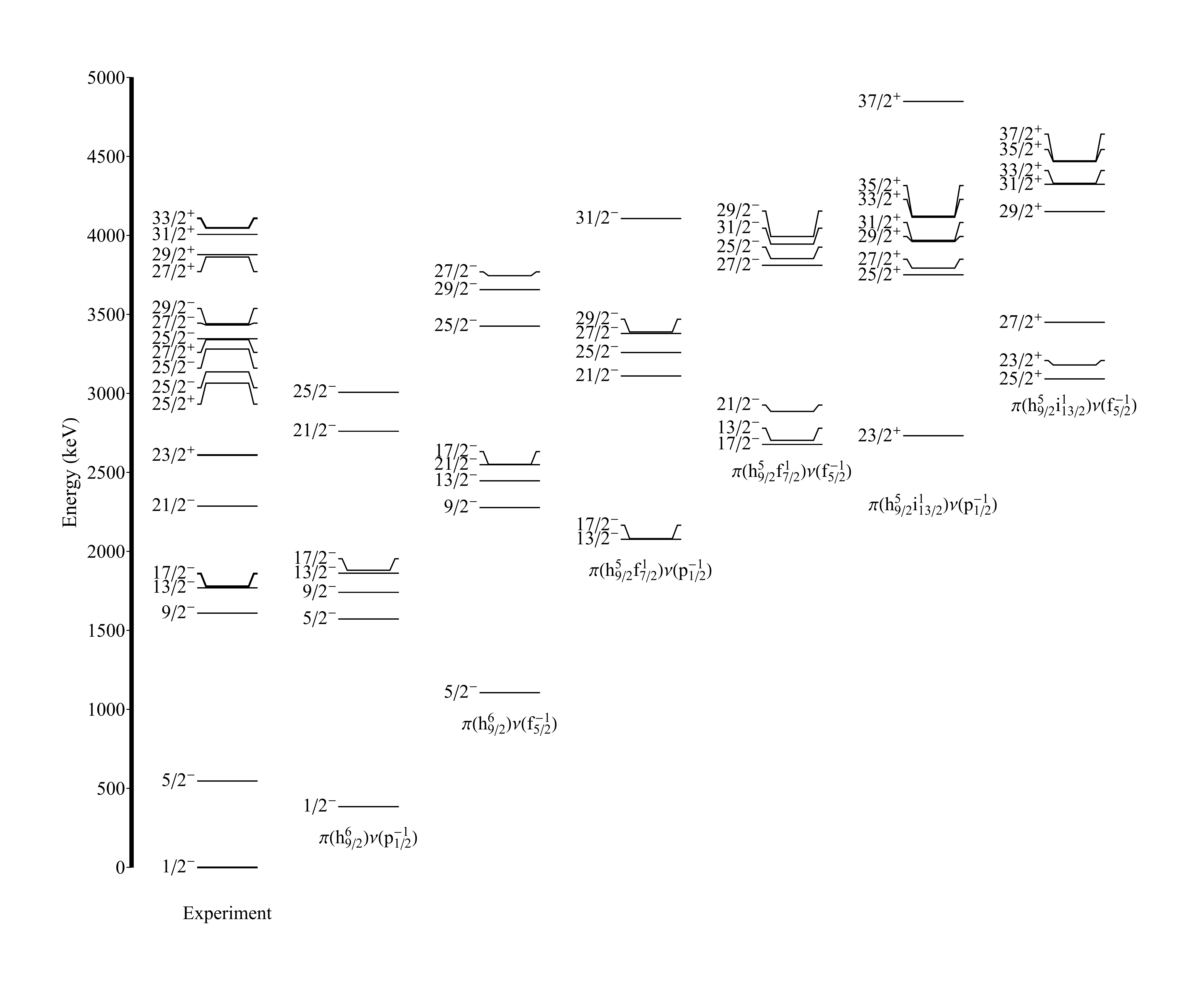}
\caption{Comparison between the experimental level scheme of $^{213}$Ra and predictions from the semiempirical shell-model calculations described in the text. Experimental data are shown on the left, while predictions from selected valence particle configurations are shown to the right. The calculated levels for $\pi[(h_{9/2}^5) i_{13/2}]\otimes \nu p_{1/2}^{-1}$ have been shifted up by 250 keV; see text.
}
\label{ra213calculation}
\end{center}
\end{figure*}

The calculated levels for $^{213}$Ra are shown in Fig.~\ref{ra213calculation} and the assigned configurations are listed in Table~\ref{table:ra213configs}. The experimental energies of the low-excitation, low-seniority states are overestimated in the calculations, as is expected when configuration mixing is neglected. Overall, however, there are many cases in which the predictions deviate from experiment by only tens of keV, allowing configuration assignments with some confidence. In most cases where alternative configurations are available, that lying closest in energy to the experimental level is expected to be dominant in the wavefunction. As will be discussed in greater detail below, the $\pi (h_{9/2}^5 i_{13/2}) \otimes \nu p_{1/2}^{-1}$ configuration consistently under predicted the experimental energies, and the states associated with this configuration have been moved up in energy by 250 keV throughout the following discussion, including in the Figures and Tables. \\

\begin{table*}[ht]
\centering
\caption{Configuration assignments in $^{213}$Ra.}
\label{table:ra213configs}
\begin{ruledtabular}
\begin{tabular}{ccrcc}
$E_x$ (keV)	&	$J^{\pi}$	&	Nominal Configuration &$E_{\rm calc}$ (keV) & $E_{\rm calc} - E_x$ (keV)	\\

\hline
0                &    $1/2^-$    & $\pi (h_{9/2}^6)_{0^+} \otimes \nu p_{1/2}^{-1}$ &385 &385  \\
546              &    $5/2^-$    & $\pi (h_{9/2}^6)_{0^+} \otimes \nu f_{5/2}^{-1}$ &1105 &559 \\
1609             &    $9/2^-$    & $\pi (h_{9/2}^6)_{4^+} \otimes \nu p_{1/2}^{-1}$ &1741 &132  \\
1770             &   $13/2^-$    & $\pi (h_{9/2}^6)_{6^+} \otimes \nu p_{1/2}^{-1}$ &1862 &92 \\
$1770+\Delta$    &   $17/2^-$    & $\pi (h_{9/2}^6)_{8^+} \otimes \nu p_{1/2}^{-1}$ &1881 &$111-\Delta$ \\
$2287+\Delta$    &   $21/2^-$    & $\pi (h_{9/2}^6)_{8^+} \otimes \nu f_{5/2}^{-1}$ &2548 &$261-\Delta$ \\
$2610+\Delta$    &   $23/2^+$    & $\pi [(h_{9/2}^5)_{9/2} i_{13/2}]_{11^-} \otimes \nu p_{1/2}^{-1}$ &2732 \footnotemark[1] &$122-\Delta$  \\
$3065+\Delta$    &   $25/2^+$    & $\pi [(h_{9/2}^5)_{9/2} i_{13/2}]_{10^-} \otimes \nu f_{5/2}^{-1}$ &3092 &$-27-\Delta$ \\
$3136+\Delta$    &   $25/2^-$    & $\pi (h_{9/2}^6)_{12^+} \otimes \nu p_{1/2}^{-1}$  &3008 &$-128-\Delta$ \\
$3281+\Delta$    &   $25/2^-$    & $\pi [(h_{9/2}^5)_{17/2} f_{7/2}]_{12^+} \otimes \nu p_{1/2}^{-1}$ &3259 &$-22-\Delta$\\
$3340+\Delta$    &   $27/2^+$    & $\pi [(h_{9/2}^5)_{9/2} i_{13/2}]_{11^-} \otimes \nu f_{5/2}^{-1}$ &3590 &$110-\Delta$  \\
$3345+\Delta$    &   $25/2^-$    & $\pi (h_{9/2}^6)_{10^+} \otimes \nu f_{5/2}^{-1}$ &3426 &$81-\Delta$ \\
$3433+\Delta$    &   $27/2^-$    & $\pi [(h_{9/2}^5)_{21/2} f_{7/2}]_{13^+} \otimes \nu p_{1/2}^{-1}$ &3379 &$-54-\Delta$ \\
$3441+\Delta$    &   $29/2^-$    & $\pi [(h_{9/2}^5)_{21/2} f_{7/2}]_{14^+} \otimes \nu p_{1/2}^{-1}$ &3389 &$-52-\Delta$ \\
$3863+\Delta$    &   $27/2^+$    & $\pi [(h_{9/2}^5)_{13/2} i_{13/2}]_{13^-} \otimes \nu p_{1/2}^{-1}$ &3793 \footnotemark[1] &$-70-\Delta$  \\
$3878+\Delta$    &   $29/2^+$    & $\pi [(h_{9/2}^5)_{17/2} i_{13/2}]_{14^-} \otimes \nu p_{1/2}^{-1}$ &3961 \footnotemark[1]&$83-\Delta$ \\
$4007+\Delta$    &   $31/2^+$    & $\pi [(h_{9/2}^5)_{17/2} i_{13/2}]_{15^-} \otimes \nu p_{1/2}^{-1}$ &3969 \footnotemark[1]&$-38-\Delta$ \\
$4048+\Delta$    &   $33/2^+$    & $\pi [(h_{9/2}^5)_{21/2} i_{13/2}]_{16^-} \otimes \nu p_{1/2}^{-1}$ &4115 \footnotemark[1]&$67-\Delta$ \\
$-$    &   $35/2^+$    & $\pi [(h_{9/2}^5)_{21/2} i_{13/2}]_{16^-} \otimes \nu p_{1/2}^{-1}$ &4122 \footnotemark[1] &$-$ \\
$-$    &   $37/2^+$    & $\pi [(h_{9/2}^5)_{21/2} i_{13/2}]_{16^-} \otimes \nu p_{1/2}^{-1}$ &4848 \footnotemark[1] &$-$ \\
$-$    &   $35/2^+$    & $\pi [(h_{9/2}^5)_{21/2} i_{13/2}]_{16^-} \otimes \nu f_{5/2}^{-1}$ &4467 &$-$ \\
$-$    &   $37/2^+$    & $\pi [(h_{9/2}^5)_{21/2} i_{13/2}]_{16^-} \otimes \nu f_{5/2}^{-1}$ &4472 &$-$ \\
\end{tabular}
\end{ruledtabular}
\footnotetext[1]{Calculated energy of this state has been increased by 250 keV as discussed in the text.}
\end{table*}

\noindent
\textbf{\textit{Negative-parity configurations}} \\

The following configurations account for the observed negative-parity states in $^{213}$Ra:

\begin{itemize}
\item $\pi h_{9/2}^6 \otimes \nu p_{1/2}^{-1}$; \\[-0.5cm]
\item $\pi h_{9/2}^6 \otimes \nu f_{5/2}^{-1}$; and \\[-0.5cm]
\item $\pi (h_{9/2}^5 f_{7/2}^1) \otimes \nu p_{1/2}^{-1}$.
\end{itemize}

The $\pi h_{9/2}^6~\otimes~\nu p_{1/2}^{-1}$ configuration can produce states up to a maximum $J^{\pi}$~=~25/2$^-$. As expected, the low-seniority states associated with this configuration and $\pi h_{9/2}^6~\otimes~\nu f_{5/2}^{-1}$ are poorly reproduced as the effects of configuration mixing are not considered \cite{rn212,ra214,ra215}. For instance, the energy of the ground state, where additional (pairing) correlations are present,  is overestimated by 384~keV. However, the calculations do reproduce the energy separation of the first-excited $J^{\pi}$~=~5/2$^-$ state and the ground state, confirming the conclusion of the weak-coupling calculation that the $J^{\pi}$~=~5/2$^-$ state is predominantly due to the movement of the neutron hole from the $\nu p_{1/2}^{-1}$ orbital to the $\nu f_{5/2}^{-1}$ orbital. At higher spin (and higher seniority for $\pi h_{9/2}^6$) the calculated energies are closer to experiment. The energy separation between the $J^{\pi}$~=~$17/2^-$, $1770+\Delta$-keV state and the $J^{\pi}$~=~$21/2^-$, $2287+\Delta$-keV state is well described and these states are associated with $\pi(h_{9/2}^6)_{8^+} \otimes \nu p_{1/2}^{-1}$ and $\pi(h_{9/2}^6)_{8^+} \otimes \nu f_{5/2}^{-1}$, respectively. Thus the 518-keV $J^{\pi}$~=~$21/2^- \rightarrow 17/2^-$ transition is analogous to the 546-keV $J^{\pi}$~=~$5/2^- \rightarrow 1/2^-$ transition.

The calculation accounts for the existence of the three $J^{\pi}$~=~25/2$^-$ states near 3.2 MeV excitation energy. Configuration mixing must be expected, however the order in the calculated level scheme is in reasonable agreement with experiment, suggesting dominant configurations, in order of excitation energy, of $\pi h_{9/2}^6~\otimes~\nu p_{1/2}^{-1}$,  $\pi (h_{9/2}^5 f_{7/2})~\otimes~\nu p_{1/2}^{-1}$, and $\pi h_{9/2}^6~\otimes~\nu f_{5/2}^{-1}$. All three states decay by $E2$ transitions to the $J^\pi$~=~21/2$^-$ state, which is expected to have significant $\pi(h_{9/2}^6)_{10^+} \otimes \nu p_{1/2}^{-1}$ admixtures along with the leading term $\pi(h_{9/2}^6)_{8^+} \otimes \nu f_{5/2}^{-1}$. Thus the $E2$ transitions can be attributed mainly to
$\pi(h_{9/2}^6)_{10^+} \otimes \nu f_{5/2}^{-1} \rightarrow \pi(h_{9/2}^6)_{8^+} \otimes \nu f_{5/2}^{-1}$ and
$\pi(h_{9/2}^6)_{12^+} \otimes \nu p_{1/2}^{-1} \rightarrow \pi(h_{9/2}^6)_{10^+} \otimes \nu p_{1/2}^{-1}$ components.

The $J^{\pi}=27/2^-$ and $J^{\pi}$~=~$29/2^-$ states at $3433 + \Delta$ keV and $3441 + \Delta$ keV have configurations that differ only by recoupling the proton spin to $\pi(h_{9/2}^5 f_{7/2})_{13^+} \otimes \nu p_{1/2}^{-1}$  and $\pi(h_{9/2}^5 f_{7/2})_{14^+} \otimes \nu p_{1/2}^{-1}$, respectively. In this scenario, the small energy spacing between the $J^{\pi}$~=~27/2$^-$ and $J^{\pi}$~=~29/2$^-$ states is accounted for. The $J^{\pi}$~=~$29/2^- \rightarrow 27/2^-$ $M1$ transition can be associated with $\pi(h_{9/2}^5 f_{7/2})_{14^+} \otimes \nu p_{1/2}^{-1} \rightarrow \pi(h_{9/2}^5 f_{7/2})_{13^+} \otimes \nu p_{1/2}^{-1}$, with a strength proportional to $(g_{h_{9/2}}-g_{f_{7/2}})^2$ where $g_{h_{9/2}}$ and $g_{f_{7/2}}$ are the $g$~factors of the $\pi h_{9/2}$ and $\pi f_{7/2}$ orbits, respectively \cite{LawsonP304,MorinagaP148}. In this case, and for similar cases in neighbouring Ra and Rn isotopes \cite{rn210,ra214}, the transition strength is of the order of 0.1 W.u., which explains the lack of an observable lifetime for the 8-keV transition.




To sum up, the negative-parity states observed in $^{213}$Ra can be attributed to the $\pi h_{9/2}^6 \otimes \nu p_{1/2}^{-1}$, $\pi h_{9/2}^6 \otimes \nu f_{5/2}^{-1}$ and $\pi(h_{9/2}^5 f_{7/2}) \otimes \nu p_{1/2}^{-1}$ configurations. \\


\noindent
\textbf{\textit{Positive-parity configurations}} \\

Two configurations giving rise to positive-parity states were considered:

\begin{itemize}
\item $\pi(h_{9/2}^5 i_{13/2}) \otimes \nu p_{1/2}^{-1}$; and \\[-0.5cm]
\item $\pi(h_{9/2}^5 i_{13/2}) \otimes \nu f_{5/2}^{-1}$.
\end{itemize}

The lowest-energy, positive-parity state observed in $^{213}$Ra, the $J^{\pi}$~=~23/2$^+$ isomer, is analogous to the $J^{\pi}$~=~11$^-$ isomer in $^{214}$Ra with the additional coupling of a $\nu p_{1/2}^{-1}$ neutron hole. The dominant configuration is $\pi(h_{9/2}^5 i_{13/2})_{11^-}\otimes \nu p_{1/2}^{-1}$. In most even-even nuclei in this region, the $J^{\pi}$~=~11$^-$ state decays via one or more $E3$ branches to $J^\pi$~=~8$^+$ states. In $^{213}$Ra, no $E3$ transition to the $J^{\pi}$~=~17/2$^-$ state was observed. This non-observation of the $E3$ decay is evidently due, in part, to the intermediate $J^{\pi}$~=~21/2$^-$ state, which allows a decay via the 323-keV $E1$ transition. In the neighboring nucleus, $^{211}$Rn, the corresponding $E1$ and $E3$ decays of the $J^\pi=23/2^+$ state were observed with a branching ratio of $I(E3)/I(E1) = (3 \pm 1)\% $. Scaling by the transition energies, the expected intensity of the $E3$ branch in $^{213}$Ra would be $\sim 2\%$. The experimental limit on the intensity of the expected 840 keV $E3$ transition is  $<2.1\%$, relative to the 323-keV $E1$ transition intensity. Thus, the expected $E3$ intensity is at our detection limit.


Moving up the level scheme, the semiempirical shell-model calculations predict the occurrence of the $J^{\pi}$~=~25/2$^+$ state (which was missed in the simplified calculations) and place it close to the experimentally observed energy of $3065+\Delta$ keV. The level arises predominantly from the coupling of the $\nu f_{5/2}^{-1}$ neutron hole to valence protons in the $\pi(h_{9/2}^5 i_{13/2})_{10^+}$ configuration.

The interpretation of the higher-spin positive-parity levels became relatively straight forward once it was recognized that the semiempirical shell model calculations for the $\pi(h_{9/2}^5 i_{13/2}) \otimes \nu p_{1/2}^{-1}$ configuration are underestimating the level energies by about 250 keV. Once the calculations are adjusted upwards by this amount, the predicted and experimental levels come into good agreement. Figure~\ref{ra213calculation} and Table \ref{table:ra213configs} show the calculated states from the $\pi[(h_{9/2}^5) i_{13/2}]\otimes \nu p_{1/2}^{-1}$ configuration with this shift included. It thus becomes evident that the lower $J^{\pi}$~=~27/2$^+$ state is associated with a dominant configuration of $\pi[(h_{9/2}^5)_{9/2} i_{13/2}]_{11}\otimes \nu f_{5/2}^{-1}$, and the upper one with $\pi[(h_{9/2}^5)_{13/2} i_{13/2}]_{13}\otimes \nu p_{1/2}^{-1}$.
The $J^{\pi}$~=~27/2$^+$ states are no doubt mixed, so decays of the $J^{\pi}$~=~27/2$^+$ states to the $J^{\pi}$~=~25/2$^+$ state, which is predominantly $\pi(h_{9/2}^5)_{11/2} i_{13/2}~\otimes~\nu f_{5/2}^{-1}$, are likely originating from $M1$ transitions between alternative spin couplings of this configuration in the initial and final states. Such transitions are relatively strong, which may explain why this part of the level scheme is decoupled from the neighboring states.



A single $J^{\pi}$~=~29/2$^+$ state is observed that can be associated with a dominant $\pi[(h_{9/2}^5)_{17/2} i_{13/2}]_{14} \otimes \nu p_{1/2}^{-1}$ configuration.
An $E1$ decay to the $J^{\pi}$~=~29/2$^-$ level can compete with $M1$ and/or $E2$ decays to the positive parity states, which for the dominant configurations are either forbidden (as is the case for the first $J^{\pi}$~=~27/2$^+$ state at $3340+\Delta$~keV) or suppressed by low transition energy and mid-shell cancellation between states of the seniority-two $\pi(h_{9/2}^5)$ configuration (as is the case for the second $J^{\pi}$~=~27/2$^+$ state at $3863+\Delta$~keV) \cite{midshell}.
The $J^\pi$~=~31/2$^+$ state with dominant configuration $\pi[(h_{9/2}^5)_{17/2} i_{13/2}]_{15} \otimes \nu p_{1/2}^{-1}$ is predicted just above the $J^{\pi}$~=~29/2$^+$ state, in agreement with experiment.

The isomeric $J^{\pi}$~=~33/2$^+$ level is also associated with the same configuration, with dominant coupling $\pi[(h_{9/2}^5)_{21/2} i_{13/2}]_{16} \otimes \nu p_{1/2}^{-1}$. The retarded 170-keV $E2$ transition to the $J^{\pi}$~=~29/2$^+$ state is thus associated with $\pi[(h_{9/2}^5)_{21/2} \rightarrow \pi[(h_{9/2}^5)_{17/2}$, which is inhibited (0.005 W.u. in $^{213}$Ra) because it represents a transition between states of the same seniority with a half-filled orbit \cite{midshell}. The equivalent decay in $^{213}$Fr has a strength of $\sim 0.05$~W.u.

The 41-keV $M1$ transition to the $J^{\pi}$~=~31/2$^+$ state can be attributed to a small $\pi[(h_{9/2}^5)_{21/2} i_{13/2}]_{15} \otimes \nu p_{1/2}^{-1}$ component in the $J^{\pi}$~=~31/2$^+$ level; the  $\pi[(h_{9/2}^5)_{21/2} i_{13/2}]_{16} \rightarrow \pi[(h_{9/2}^5)_{21/2} i_{13/2}]_{15}$ transition has a strength of about 0.1 W.u., so only small admixtures are required to account for the observed transition strength of $\sim 10^{-4}$ W.u.

A 606-keV transition from the $J^{\pi}$~=~33/2$^+$ isomer to the $J^{\pi}$~=~29/2$^-$ level is also observed, which can have mixed $M2+E3$ multipolarity. We have shown the experimental transition strengths for the extremes of pure $M2$ and pure $E3$ multipolarity in Table~\ref{transitionstrengths}.  Assuming pure multipolarity, the $M2$ strength is very weak whereas the $E3$ is unrealistically strong. The nominal configuration change, $\pi[(h_{9/2}^5)_{21/2} i_{13/2}]_{16} \otimes \nu p_{1/2}^{-1} \rightarrow \pi[(h_{9/2}^5)_{21/2} f_{7/2}]_{14} \otimes \nu p_{1/2}^{-1}$ does not permit an $M2$ transition, but does allow an $E3$ transition. The expected $E3$ strength can be scaled from the parent $\pi[(h_{9/2}^5)_{21/2} i_{13/2}]_{17^-} \rightarrow \pi[(h_{9/2}^5)_{21/2} i_{13/2}]_{14^+}$ transition in $^{214}$Ra to give an expected strength of $\sim$~10~W.u. (The scale factor is given by the ratio of Racah coefficients, $[W(\frac{21}{2} \frac{13}{2} 14~3;16 \frac{7}{2}) / W(\frac{21}{2} \frac{13}{2} 14~3;17 \frac{7}{2})]^2$.) Thus an observed transition strength of 36 W.u. for a pure $E3$ transition (Table~\ref{transitionstrengths}) is unrealistic, and an $M2$ component must be present.

An $M2$ component in the 606-keV transition could be explained by small mixing of the configuration $\pi[h_{9/2}^4 f_{7/2} i_{13/2}]_{16^-} \otimes \nu p_{1/2}^{-1} $ in the $J^\pi$~=~33/2$^+$ level. Although this configuration is not needed to explain the yrast spectroscopy of $^{213}$Ra, the parent proton configurations  $\pi[h_{9/2}^5 i_{13/2}]$ and $\pi[h_{9/2}^4 f_{7/2} i_{13/2}]$ are identified in $^{214}$Ra. Only a small contribution is needed because the allowed $M2$ transition  $\pi i_{13/2} \rightarrow \pi h_{9/2}$ has a strength of the order of 1~W.u. \cite{RobertsPhysRevC.93.014309}.


Theory and experiment can be brought into harmony if the 606-keV transition has mixed multipolarity with mixing ratio $|\delta(E3/M2)| \sim 0.7$.

It is not clear why the upward shift of $\sim 250$~keV is necessary for the $\pi[(h_{9/2}^5) i_{13/2}]\otimes \nu p_{1/2}^{-1}$ configuration. However, it has to be kept in mind that the semiempirical shell model calculations do not include configuration mixing. Another consideration is that the empirical single-particle energies and two-body residual interactions include octupole-coupled components due to the octupole vibration of the $^{208}$Pb core \cite{HAMAMOTO1974}. Such components could be double counted in the evaluation of more complex configurations, or affected by Pauli-blocking. Overall, given their simplicity, the agreement of the semiempirical single-configuration calculations with experiment at the level of a few hundred keV is rather remarkable.

A number of additional $\gamma$-ray transitions were observed in the experiment that must originate from higher-excited states in $^{213}$Ra, but we were not able to place them firmly in the level scheme. The calculations predict a $J^{\pi}$~=~$35/2^+$ state within 20~keV of the $J^{\pi}$~=~$33/2^+$ state for which we could not find firm experimental evidence. An $M1$ transition between the proposed configurations of the $J^{\pi}$~=~$33/2^+$ and $J^{\pi}$~=~$35/2^+$ states would not be isomeric unless its energy was less than about 20~keV. The strongest observed line above the $J^{\pi}$~=~$33/2^+$ isomer has an energy of 459-keV and is likely an $M1$ transition. Should this transition be placed above the proposed but unobserved low-energy $M1$ transition between the $J^{\pi}$~=~$35/2^+$ and $J^{\pi}$~=~$33/2^+$ states, it would depopulate a $J^{\pi}$~=~$37/2^+$ state at an excitation energy of about 4.5~MeV, for which there is a predicted candidate state with configuration $\pi[(h_{9/2}^5)_{21/2} i_{13/2}]_{17} \otimes \nu f_{5/2}^{-1}$ at an energy of 4472 keV.

In summary, we are able to assign dominant configurations to the observed states in $^{213}$Ra up to the $J^{\pi}$~=~$33/2^+$ isomer, obtaining reasonable agreement with the level energies and at least a qualitative understanding of the observed electromagnetic transitions.

\begin{figure*}[]
\begin{center}
\includegraphics[width=17cm]{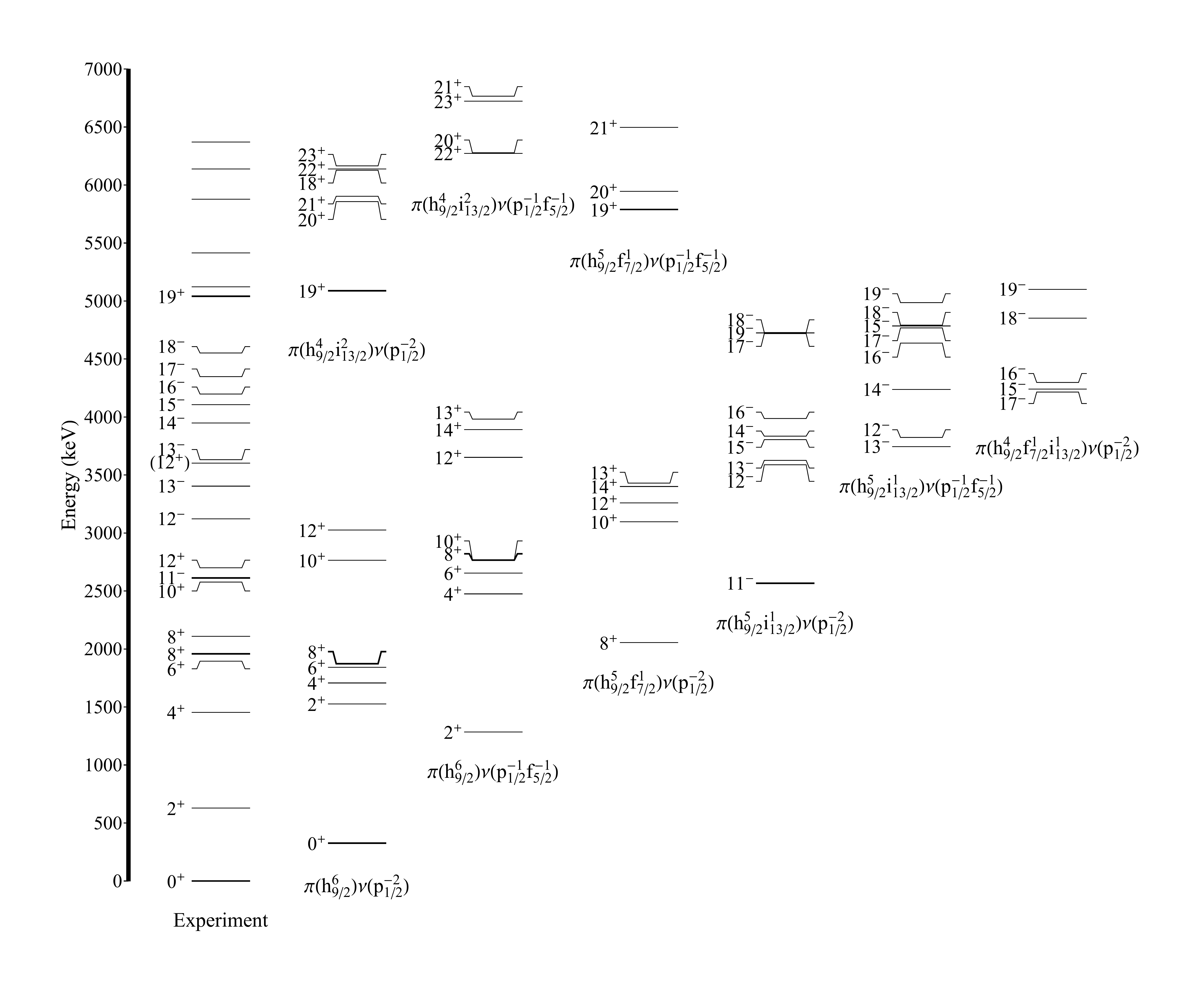}
\caption{Comparison between the experimental level scheme of $^{212}$Ra and predictions from the semiempirical shell-model calculations described in the text. Experimental data are shown on the left, while predictions for selected configurations are shown to the right.  For the $J > 19$ levels, no spin or parity assignments have been made.}
\label{ra212calculation}
\end{center}
\end{figure*}




\subsubsection{\textbf{$^{212}$Ra Calculation}}

At the level of the weak-coupling model discussed above in relation to $^{213}$Ra, an equivalent assumption for $^{212}$Ra is that the two neutron holes couple to spin-parity $J^{\pi}$~=~0$^+$ such that the level scheme of $^{212}$Ra then should resemble that of $^{214}$Ra. A weak-coupling model calculation along these lines was performed by Kohno \textit{et al.}~\cite{ra212}. A comparison of the level schemes of $^{212}$Ra and $^{214}$Ra suggests equivalence of the following yrast states: $J^{\pi}$~=~0$^+$, $J^{\pi}$~=~6$^+$, $J^{\pi}$~=~8$^+$, $J^{\pi}$~=~11$^-$, and $J^{\pi}$~=~17$^-$, along with the second $J^{\pi}$~=~8$^+$ state. Otherwise, more detailed conclusions cannot be drawn.  In fact, when it comes to the semiempirical shell model calculations, the two neutron holes in $^{212}$Ra open up a larger array of available configurations, including two-proton and two-neutron excitations from the lowest energy configuration, $\pi h_{9/2}^6 \otimes \nu p_{1/2}^{-2}$. Of the many possible combinations that were examined, only the nine most relevant to the observed states will be discussed. These are: \\[-0.5cm]

\begin{itemize}
\item $\pi h_{9/2}^6 \otimes \nu p_{1/2}^{-2}$
\item $\pi h_{9/2}^6 \otimes \nu(p_{1/2}^{-1}f_{5/2}^{-1})$
\item $\pi (h_{9/2}^5 f_{7/2}) \otimes \nu p_{1/2}^{-2}$
\item $\pi (h_{9/2}^5 f_{7/2}) \otimes \nu (p_{1/2}^{-1}f_{5/2}^{-1})$  
\item $\pi (h_{9/2}^5i_{13/2}) \otimes \nu p_{1/2}^{-2}$  
\item $\pi (h_{9/2}^5i_{13/2})\otimes \nu(p_{1/2}^{-1}f_{5/2}^{-1})$
\item $\pi (h_{9/2}^4f_{7/2} i_{13/2}) \otimes \nu p_{1/2}^{-2}$ 
\item $\pi (h_{9/2}^4i_{13/2}^2) \otimes \nu p_{1/2}^{-2}$
\item $\pi (h_{9/2}^4i_{13/2}^2) \otimes \nu (p_{1/2}^{-1}f_{5/2}^{-1})$ 
\end{itemize}

The semiempirical shell model calculations for $^{212}$Rn are compared with experiment in Fig.~\ref{ra212calculation}.
As usual, the energies of the low-seniority, low-excitation states, particularly the ground state and first-excited $J^{\pi}$~=~2$^+$ state, are over-estimated by the calculation. The ground-state configuration is nominally $\pi h_{9/2}^6 \otimes \nu p_{1/2}^{-2}$. Like in  $^{213}$Ra, the first-excited state is associated with the excitation of a neutron hole, $\nu p^{-1}_{1/2}$~$\rightarrow$~$\nu f^{-1}_{5/2}$. However there is evidently considerable configuration mixing as the lowest calculated $J^{\pi}$~=~2$^+$ state, associated with $\pi h_{9/2}^6 \otimes \nu(p_{1/2}^{-1}f_{5/2}^{-1})$, is approximately 600~keV higher than the experimental energy.


By contrast, the $J^{\pi}$~=~6$^+$ and $J^{\pi}$~=~8$^+$ states are well explained by the pure $\pi h_{9/2}^6 \otimes \nu(p_{1/2}^{-2})$ configuration. The second $J^{\pi}$~=~8$^+$ state is well explained by the pure $\pi(h_{9/2}^5 f_{7/2})\otimes \nu p_{1/2}^{-2}$ configuration, with the same $\pi f_{7/2}$ proton excitation creating the corresponding state in $^{214}$Ra \cite{ra214}.

The $J^{\pi}$~=~11$^-$ state in $^{212}$Ra is produced by the same $\pi i_{13/2}$ proton excitation as in $^{214}$Ra, and is also analogous to the $J^{\pi}$~=~23/2$^+$ yrast state in $^{213}$Ra. In each of these isotopes the related states lie at an excitation energy of approximately 2600~keV and are isomeric due to the change in configuration, which includes a parity change. In $^{214}$Ra, two $E3$ transitions to the $J^\pi$~=~8$^+$ states carry 100$\%$ of the decay intensity from the $J^{\pi}$~=~11$^-$ state, while in $^{212}$Ra a 36-keV $E1$ transition to the $J^{\pi}$~=~10$^+$ state is the dominant branch, with the two $E3$ transitions to the $J^{\pi}$~=~8$^+$ states sharing half of the decay intensity. The transition strengths in $^{214}$Ra are 3.1(1)~W.u for the nominal $\pi(h_{9/2}^5 i_{13/2}$)~$\rightarrow$~$h_{9/2}^6$ transition and 21.7(6)~W.u. for the nominal $\pi(h_{9/2}^5 i_{13/2}$)~$\rightarrow$~$\pi(h_{9/2}^5 f_{7/2}$) transition. In $^{212}$Ra, the $E3$ strengths have decreased slightly but their proportionality has remained the same, namely 2.1(6)~W.u for $\pi i_{13/2}$ $\rightarrow$ $\pi h_{9/2}$ and 14.4(3)~W.u. for $\pi i_{13/2}$ $\rightarrow$ $\pi f_{7/2}$. This similarity between $^{214}$Ra and $^{212}$Ra supports the configuration assignments for the two $J^{\pi}$~=~8$^+$ states and the $J^{\pi}$~=~11$^-$ isomer.

Compared with $^{214}$Ra, the $J^{\pi}$~=~10$^+$ state in $^{212}$Ra has been driven below the $J^{\pi}$~=~11$^-$ state; the same behavior has been observed in the $N = 124$ isotone, $^{210}$Rn \cite{rn210}. The calculations predict three $J^{\pi}$~=~10$^+$ states due to the $\pi h_{9/2}^6 \otimes \nu p_{1/2}^{-2}$, $\pi h_{9/2}^6 \otimes \nu(p_{1/2}^{-1}f_{5/2}^{-1})$ and $\pi(h_{9/2}^5f_{7/2}) \otimes \nu p_{1/2}^{-2}$ configurations, all higher in energy than the $J^{\pi}$~=~11$^-$ state. However, if mixing were taken into account, it can be expected that the yrast $J^{\pi}$~=~10$^+$ state would be pushed down in energy and could come below the $J^{\pi}$~=~11$^-$ level, as is observed.

Above the $J^{\pi}$~=~11$^-$ state, it becomes difficult to match the semiempirical shell-model configurations with the observed excited states. Nevertheless, the experimental states between $J^{\pi}$~=~11$^-$ and $J^{\pi}$~=~18$^-$ appear to be associated with three dominant negative-parity configurations, $\pi (h_{9/2}^5i_{13/2}) \otimes \nu p_{1/2}^{-2}$, $\pi (h_{9/2}^5i_{13/2})\otimes \nu(p_{1/2}^{-1}f_{5/2}^{-1})$, and $\pi (h_{9/2}^4f_{7/2} i_{13/2}) \otimes \nu p_{1/2}^{-2}$. The high level density implies that a high level of configuration mixing may be required to describe this portion of the level scheme. The incremental sequence of negative-parity states from $J^{\pi}$~=~13$^-$ to $J^{\pi}$~=~18$^-$ means that the $J^{\pi}$~=~17$^-$ state, which is isomeric in $^{214}$Ra, is no longer an isomer in $^{212}$Ra.
More specifically, in $^{214}$Ra, the yrast $J^{\pi}$~=~17$^-$ state decays via an enhanced $E3$ transition to the yrast $J^{\pi}$~=~14$^+$ state. The $J^{\pi}$~=~17$^-$ level is nominally associated with the $\pi (h_{9/2}^5i_{13/2})$ configuration and includes mixing with $\pi (h_{9/2}^4f_{7/2} i_{13/2})$, which can be included using the multiparticle octupole coupling model approach \cite{MPOC-POLETTI1986,rn212,ra214}. In $^{212}$Ra, however, the yrast $J^{\pi}$~=~14$^+$ state is not observed and the $J^{\pi}$~=~17$^-$ state is not isomeric, decaying via an $M1$ transition to $J^{\pi}$~=~16$^-$.

The $J^\pi$~=~19$^+$ isomer can be identified with the $\pi (h_{9/2}^4i_{13/2}^2)\otimes \nu(p_{1/2}^{-2})$ configuration. It decays via an $E1$ transition to the $J^{\pi}$~=~18$^-$ state with a typical $E1$ strength of 6.7(8)$\times$10$^{-8}$ W.u.. A search was made for possible $E3$ transitions from this state, but none could be found. In most cases, the strong $E3$ transitions take place between the maximal spin couplings of the initial and final configurations, whereas, in this case in $^{212}$Ra, the angular momentum coupling of the initial and final states inhibits $E3$ transitions.

It is noteworthy that the corresponding $J^\pi$~=~19$^+$ state is not observed in $^{214}$Ra, and instead the $J^{\pi}$~=~18$^+$ state of the $\pi (h_{9/2}^4i_{13/2}^2)$ configuration is observed. The addition of the two neutron holes leads to a considerable re-ordering of the spin sequence within the proton configuration such that the $J^{\pi}$~=~18$^+$ level does not appear in the experimental level scheme of $^{212}$Ra. The calculation predicts that it is pushed above the $J^{\pi}$~=~21$^+$ state. Such relatively small changes in the level ordering can have a profound influence on the experimentally accessible near-yrast level sequence and decay scheme.

Several states lying above $J^{\pi}$~=~19$^+$ were observed but have no firm spin-parity assignments. \\


\section{Conclusions}

The radium isotopes with one and two neutron holes in the $N=126$ closed shell have been investigated experimentally, $^{213}$Ra to spins of about 33/2$\hbar$ and excitation energies of about 4 MeV, and $^{212}$Ra to spins of about 20$\hbar$ and excitation energies of about 6 MeV.
Knowledge of $^{213}$Ra has been extended beyond the $J^{\pi}$~=~17/2$^-$, $\tau$=3-ms metastable state. Two new isomers with $J^{\pi}$~=~23/2$^+$, $\tau$~=~27(3)~ns and $J^{\pi}$~=~33/2$^+$, $\tau$ = 50(3)~ns have been identified. In $^{212}$Ra, an isomer with $J^{\pi}$~=~19$^+$ and $\tau$~=~31(3)~ns has been identified.

Semiempirical shell-model calculations give a satisfactory description of the excitation energies and decay rates in $^{213}$Ra up to the $J^{\pi}$~=~33/2$^+$ isomer. A reasonable account can also be given for $^{212}$Ra. Large-basis shell-model calculations are now feasible in many regions of the nuclear chart, however the region near $^{208}$Pb has been largely neglected to date; it would be timely to re-visit this region. The success of the semiempirical shell-model calculations suggests that deeper insights into the evolution of nuclear structure and the onset of collectivity could be gained by developing a more sophisticated theory that includes configuration interactions.

\section*{Acknowledgments}

The authors are grateful to the academic and technical staff of the Department of Nuclear Physics (Australian National University) and the Heavy Ion Accelerator Facility for their continued support.
This research was supported in part by the Australian Research Council grants numbers DP120101417, DP130104176, DP140102986, DP140103317, and FT100100991.
A.A. and M.S.M.G. acknowledge support of the Australian Government Research Training Program. Support for the ANU Heavy Ion Accelerator Facility operations through the Australian National Collaborative Research Infrastructure Strategy (NCRIS) program is acknowledged.

\bibliography{Radium}

\end{document}